\documentclass[lettersize, journal]{IEEEtran}
\usepackage{amsmath,amsfonts}
\usepackage{algorithmic}
\usepackage{algorithm}
\usepackage{array}
\usepackage[caption=false,font=footnotesize,labelfont=sf,textfont=sf]{subfig}
\usepackage{textcomp}
\usepackage{stfloats}
\usepackage{url}
\usepackage{verbatim}
\usepackage{graphicx}
\usepackage{cite}
\usepackage{flushend}
\usepackage{multicol}
\usepackage{amssymb}
\usepackage{array}
\usepackage{hhline}
\usepackage{amsmath}
\usepackage{adjustbox}
\usepackage{wrapfig}
\usepackage{booktabs} 
\usepackage{graphicx}
\usepackage{xspace}
\usepackage{mathtools}
\usepackage{soul}
\usepackage{multirow}
\usepackage{comment}
\usepackage{etoolbox}
\usepackage{xcolor}
\usepackage{eucal}
\usepackage{pifont} 
\usepackage{wrapfig}
\usepackage{threeparttable}
\usepackage{wasysym}
\usepackage{soul}
\usepackage{MnSymbol} 
\usepackage{orcidlink}
\usepackage{tikz}
\usepackage{stackengine}

\usepackage{float} 

\usepackage{enumitem}
\setlist[itemize]{leftmargin=*}

\newcommand{\crm}[1]{\textcolor{black}{#1}} 

\newcommand{\sysname}{\textsc{Touch-to-Pair}\xspace}
\newcommand{\shortname}{\textsc{T2Pair}\xspace}
\newcommand{\faithful}{\textsc{T2Pair}\xspace}

\newcommand{\precise}{\textsc{T2Pair++}\xspace}
\newcommand{\pname}{\crm{TimeWall}\xspace}

\newcommand{\eg}{e.g.,\ }
\newcommand{\ie}{i.e.,\ }
\newcommand{\etal}{\latinphrase{et~al.}\xspace}

\newcommand{\latinphrase}[1]{\textit{#1}}  

\newcommand{\cxr}[1]{\textcolor{black}{#1}} 

\newcommand{\zm}[1]{{\color{black}#1}} 

\hypersetup{
    colorlinks,
    citecolor=black,
    filecolor=black,
    linkcolor=black,
    urlcolor=black
}

\newcommand{\aaf}{\vspace*{-6pt}}
\newcommand{\af}{\vspace*{-3pt}}

\DeclareMathAlphabet{\mathpzc}{OT1}{pzc}{m}{it}

\newcommand{\cmark}{\ding{51}}%
\newcommand{\xmark}{\ding{55}}%
\newcommand{\halfcheck}{\cmark\kern-1.1ex\raisebox{.7ex}{\rotatebox[origin=c]{125}{--}}}

\newcommand*\circled[1]{\tikz[baseline=(char.base)]{
            \node[shape=circle,draw,inner sep=0.5pt] (char) {#1};}}
            
\newcolumntype{P}[1]{>{\centering\arraybackslash}p{#1}}

\usepackage{url}
\makeatletter
\g@addto@macro{\UrlBreaks}{\UrlOrds}
\makeatother

\def\BibTeX{{\rm B\kern-.05em{\sc i\kern-.025em b}\kern-.08em
T\kern-.1667em\lower.7ex\hbox{E}\kern-.125emX}}

\usepackage{balance}

\begin{document}
\title{\crm{Touch to Pair}: Secure and Usable IoT Pairing without Information Loss}

\author{Chuxiong~Wu\,\orcidlink{0000-0003-0243-661X}, 
Xiaopeng~Li\,\orcidlink{0000-0001-7686-0283},
Lannan~Luo\,\orcidlink{0000-0003-2476-7831},
and~Qiang~Zeng\,\orcidlink{0000-0001-9432-6017}%
\IEEEcompsocitemizethanks{
    \IEEEcompsocthanksitem 
    Chuxiong~Wu and Xiaopeng~Li contributed equally to this work. 
    The protocols were primarily developed by the corresponding author. 
    \IEEEcompsocthanksitem 
    Chuxiong~Wu was with the Department of Computer Science, George Mason University, Fairfax, VA 22032, USA. 
    He is now with the School of Computing, Southern Illinois University, Carbondale, IL 62901, USA. 
    Most of this work was completed while he was with George Mason University. 
    E-mail: chuxiong.wu@siu.edu.
    \IEEEcompsocthanksitem 
    Xiaopeng~Li is with the Department of Computer Science, University of Central Oklahoma, Edmond, OK 73034, USA. 
    E-mail: xli25@uco.edu.
    \IEEEcompsocthanksitem
    Lannan~Luo and Qiang~Zeng* are with the Department of Computer Science, George Mason University, Fairfax, VA 22032, USA. 
    E-mail: \{lluo4, zeng\}@gmu.edu.
    \IEEEcompsocthanksitem 
    The manuscript extends our prior work, T2Pair, published in CCS’20~\cite{li2020t2pair}. 
    T2Pair relies on fuzzy commitment, causing information loss, 
    while T2Pair++ is based on a novel pairing protocol that avoids information loss, 
    achieving higher accuracy and security.
    \IEEEcompsocthanksitem
    *Corresponding author.
}}

\markboth{}{}

\maketitle
\begin{abstract}
Secure pairing is essential for trustworthy deployment and operation of Internet of Things (IoT) devices. However, traditional pairing methods are unsuitable due to the lack of user interfaces such as keyboards.
Proximity-based approaches are usable but vulnerable to nearby attackers, while methods relying on physical operations (e.g., shaking) offer higher security but require inertial sensors that most IoT devices lack.
We introduce \emph{Universal Operation Sensing}, which enables IoT devices to detect user operations without inertial sensors.
With this technique, users can complete pairing within seconds through simple actions, such as pressing a button or twisting a knob, using either a smartphone or a smartwatch.
We further identify an accuracy issue caused by information loss in the commonly used fuzzy-commitment protocol.
To address this issue, we propose \emph{\pname}, an accurate pairing protocol that avoids fuzzy commitment and incurs \emph{zero} information loss.
A comprehensive evaluation shows that it is secure, usable, and efficient.

\end{abstract}


\section{Introduction}\label{sec:Introduction}
\IEEEPARstart{I}{nternet-of-Things} (IoT) devices are extensively deployed, significantly influencing diverse industries and everyday life. 
The global IoT market was \$970 billion in 2022 and is projected to reach \$2,227 billion in 2028~\cite{iot_market}. Since pairing is essential for establishing a protected communication channel, secure and usable pairing is crucial for the widespread adoption and trustworthy operation of IoT devices.

\zm{Simple mechanisms like long-pressing a button to put an IoT device into pairing mode are vulnerable to various attacks, as they often lack proper authentication of the peer device's identity.}
\zm{For instance, an attacker can impersonate a legitimate device during pairing~\cite{zhang17_infocom, choi2021push,sethi2019misbinding}, a problem especially acute in environments with many nearby devices (e.g., apartments, offices, or public spaces). Past studies have shown that insecure pairing can lead to unauthorized device control, data leakage, or man-in-the-middle attacks, in which an attacker impersonates the IoT device to pair with the user’s smartphone and simultaneously impersonates the smartphone to pair with the IoT device, enabling sophisticated exploitation~\cite{Hinckley_2003, shakeWell, touch_and_guard_tmc}.}

The literature offers various IoT pairing approaches, which can be categorized into two main types.
The first category establishes pairing based on device proximity, utilizing wireless signals~\cite{zhang17_infocom, ghose18_infocom, Tap-to-pair2018} or ambient context such as audio and light~\cite{2014CCS_CZP, vibration_pairing, audio_pairing}.
\zm{While these approaches are user-friendly, they remain vulnerable to impersonation attacks by co-located malicious devices.}
The second category requires physical contact or operation with the IoT device~\cite{Hinckley_2003, shakeWell, touch_and_guard_tmc}.
For instance, ShaVe/ShaCK~\cite{shakeWell} has the user hold the smartphone and IoT device together in one hand and \emph{shake} them, using the shared motion sequence for pairing.
While such methods offer enhanced security through physical operations, they are impractical because they require IoT devices to include inertial or other specialized sensors to detect user actions, and many IoT devices lack these sensors.

\begin{figure}
\centering
\includegraphics[scale=0.48, trim=180 60 2 2, clip]{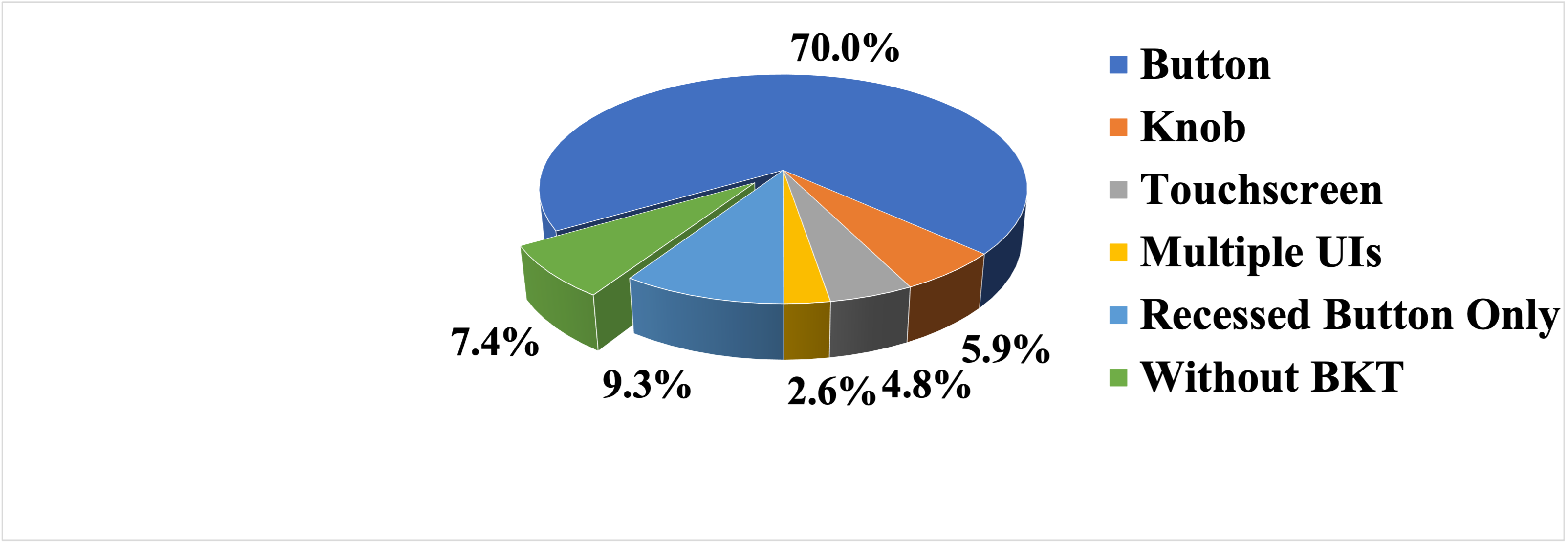}
\aaf
\caption{Distribution of UIs on 270 popular IoT devices. ``\emph{With BKT}'' means the device has
a normal Button, Knob or Touchscreen; ``\emph{Recessed button}'' refers to a small hole that
can be pressed using, e.g., a ball-point pen.}
\label{fig:UI_survey}
\aaf\aaf
\end{figure}

We consider IoT devices that (1) lack sophisticated user interfaces such as keyboards, (2) may be located near untrusted or malicious devices (for example, a hospital may contain a mix of devices belonging to doctors, patients, or even attackers), (3) do not necessarily have inertial sensors, and (4) may be mobile or stationary, installed either indoors or outdoors.
A secure and user-friendly pairing solution applicable to such heterogeneous IoT devices does not yet exist.

We introduce a novel technique,
\emph{Universal Operation Sensing (UOS)}, which allows an IoT device
to sense user operations on an IoT device without
requiring any inertial sensors. When a user wearing a wristband or holding a smartphone operates an IoT device by pressing its button, twisting its knob, or swiping its touchscreen, salient points are generated when the button is pressed/released or when the twisting/swiping direction changes. Leveraging the insight that every IoT device possesses a clock, we use timestamps to represent these salient points.
By analyzing the motion data captured by the built-in inertial measurement unit (IMU) on the user's wristband or smartphone, the same set of salient points can be identified.
Consequently, both the wristband and the IoT device can utilize the shared knowledge of these salient points for pairing.

The technique is highly user-friendly: a user wearing a wristband or holding a smartphone only needs to \emph{touch} the IoT device, through a few simple operations, to complete pairing. Moreover, the technique can be broadly applied to most IoT devices on the market. As illustrated in Figure~\ref{fig:UI_survey}, our survey of the 270 most popular IoT devices on Amazon (ranked by number of reviews) shows that 92.6\% ($=1 - 7.4\%$) feature a regular button, knob, touchscreen, or recessed button. For example, an Amazon smart plug, which requires minimal user interaction, is equipped with a button for pairing and toggling its power state.


A key difference between pairing and authentication~\cite{mare2018saw, Monrose_97, zebra, g2auth, momatch, p2auth, wu2024you, sharp2022authentication} is that the latter typically assumes that two parties have already established a security association, such as a shared key, whereas pairing does not. Thus, the pairing information cannot be directly exchanged, as it may be intercepted by an attacker. Initially, we explored the fuzzy-commitment scheme, which aims to establish a shared key while accommodating small differences in observations~\cite{Juels_FCS_CCS99}. However, our experiments with the original fuzzy-commitment scheme resulted in high inaccuracy. Our investigation revealed that minor differences between observations may lead to substantially different encodings (and hence false rejections), whereas significant differences might yield similar encodings (and hence false acceptances).

We design two protocols to address the inaccuracy issue. 
(1) A faithful fuzzy-commitment protocol \emph{partially} mitigates the issue through a novel encoding algorithm. The system implementing this protocol is named \sysname{} (\faithful{}, for short). 
(2) A \emph{\pname} pairing protocol \emph{completely} resolves the issue by introducing the notion of a ``commitment deadline,'' where any commitment sent by an attacker after the deadline is discarded. The system implementing this \pname{} pairing protocol is named \precise{}.

We implement \faithful{} and \precise{} and evaluate them on prototypical IoT devices equipped with buttons, knobs, or touchscreens. The evaluation results demonstrate high accuracy (AUC~$>0.999$ with \precise{}). Moreover, the pairing operations can be finished within \emph{four} seconds. A user study further confirms the high usability of our system. 

Our contributions are summarized as follows:

\begin{itemize}

    \item We introduce \textit{Universal Operation Sensing}, which enables IoT devices to sense user operations without relying on inertial sensors. 
This technique can therefore be applied to heterogeneous IoT devices.

\item We propose faithful fuzzy commitment that ensures differences between values are accurately reflected in the distances between their encodings, despite information loss. 
We also present a deadline-based pairing protocol, \emph{\pname}, which guarantees zero information loss during pairing.

\item Leveraging these techniques, we implement \shortname{} and \precise{}. 
A comprehensive evaluation demonstrates high accuracy, as well as strong security and usability.

\end{itemize}

The paper is organized as follows. Section~\ref{sec:system_overview} presents the system overview and threat model. Section~\ref{sec:evidence} elaborates on universal operation sensing. Section~\ref{sec:protocol} introduces the pairing protocol based on fuzzy commitment, and Section~\ref{sec:timing_protocol} the TimeWall protocol with zero information loss.
Section~\ref{sec:implementation} delves into the implementation details. Following that, Section~\ref{sec:data_collection} describes the dataset collection process, and Section~\ref{sec:evaluation} presents the evaluation results. 
The usability study is presented in Section~\ref{sec:user_study}.
We discuss related work in Section~\ref{sec:related_work} and outline the limitations in Section~\ref{sec:limitations}. Finally, we conclude our paper in Section~\ref{sec:conclusion}.

\aaf
\section{System Overview and Threat Model}
\label{sec:system_overview}

\begin{figure*}
\centering
\includegraphics[scale=0.8]{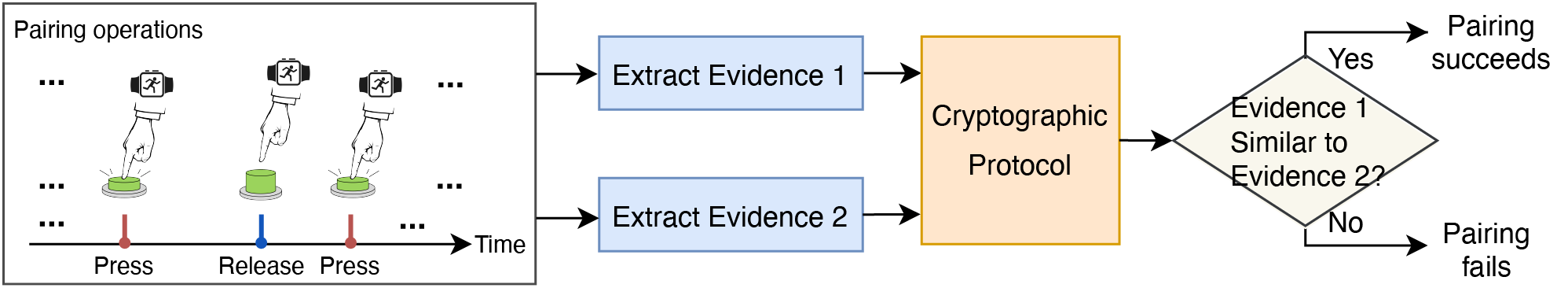}
\aaf
\caption{Architecture (a wristband as the helper and an IoT device with a button as an example).}
\label{fig:system_overview}
\aaf\af
\end{figure*}

\noindent \textbf{System Overview.} 
We illustrate the architecture of the pairing system in Figure~\ref{fig:system_overview}. 
In our design, a user employs their personal mobile device, referred to as a \emph{helper}, such as a smartphone, fitness tracker, or smartwatch, to pair an IoT device by performing a few simple operations on the device.
For example, in Figure~\ref{fig:system_overview}, a user wearing a smartwatch presses the button of an IoT device a few times. 
During this interaction, the button utilizes its internal clock to record the button-pressing events and generates a piece of \emph{evidence} for the pairing. Simultaneously, the helper device collects readings from its accelerometer/gyroscope, independently generating another piece of evidence about the operations. 
The pairing protocol allows the two parties to securely compare their evidence. If the two pieces of evidence are highly similar, the pairing succeeds and the they are able to agree on a key.

\vspace{3pt}
\noindent \textbf{Threat Model.} 
\zm{The attacker $\mathcal{A}$ may have one or more of the following goals:  
(G1) The malicious device ${D}_a$, controlled by $\mathcal{A}$, impersonates the legitimate IoT device,  
tricking the victim $\mathcal{V}$'s helper ${H}_v$ into pairing with  
${D}_a$ and sending data to it (e.g., WiFi credentials, photos, or health data)~\cite{zhang17_infocom, choi2021push,sethi2019misbinding}.  
(G2) The malicious helper ${H}_a$  impersonates the legitimate  
helper to pair with the legitimate IoT device ${D}_v$ for unauthorized device control or data access.  
(G3) The attacker cracks the key or launches a man-in-the-middle attack  by impersonating both  
the helper and the IoT device, in order to  
eavesdrop on or manipulate the data exchange between ${H}_v$ and ${D}_v$~\cite{Hinckley_2003, shakeWell, touch_and_guard_tmc}.}

Similar to prior work~\cite{Perceptio, 2014CCS_CZP}, our approach follows a standard Dolev-Yao adversary model~\cite{Dolev_Yao}, where a strong adversary has full control over all communication channels. Given this model, $\mathcal{A}$ may launch man-in-the-middle attacks, such as intercepting messages sent between ${D}_v$ and ${H}_v$ 
and substituting them with fake messages. We assume that $\mathcal{A}$ has full knowledge of our pairing protocol and thus further consider the attacks below.

\emph{\textbf{Mimicry Attacks.}} If $\mathcal{A}$ has visual observation capabilities of $\mathcal{V}$, they may launch a mimicry attack by imitating $\mathcal{V}$'s pairing operations, aiming to achieve goals G1 and/or G2. We consider various threat scenarios reflecting $\mathcal{A}$'s increasing capabilities.
\textbf{MA-obstructed:} $\mathcal{A}$ can observe $\mathcal{V}$ but is obstructed from directly seeing $\mathcal{V}$'s hand motions due to certain obstacles.
\textbf{MA-clear:} $\mathcal{A}$ can clearly observe $\mathcal{V}$'s hand motions by selecting an optimal viewing angle.
\textbf{MA-trained:} $\mathcal{A}$ is familiar with $\mathcal{V}$ and has been trained by learning $\mathcal{V}$'s pairing operations before launching a mimicry attack, as described in the MA-clear scenario.

\emph{\textbf{Brute-Force Attacks.}} \textbf{(1) BF-online:} During pairing, $\mathcal{A}$ uses a powerful computer, trying to figure out the correct evidence and  trick ${H}_v$ and/or ${D}_v$ into pairing with the attacker.
\textbf{(2)~BF-offline:} $\mathcal{A}$ gathers all pairing traffic and conducts offline analysis to crack the established key after pairing.

\textbf{\emph{Attacks beyond Scope.}} 
$\mathcal{A}$ may employ a camera and computer-vision techniques to analyze $\mathcal{V}$'s hand movements. Our system, like other pairing methods involving physical interactions such as ShaVe/ShaCK~\cite{shakeWell}, is susceptible to such attacks. However, in a user's private space, the attack is difficult to launch because it requires an attacker-controlled camera directed at the user. In public spaces, we suggest that users conceal the pairing operations with their body or the other hand, similar to shielding a keypad when entering a PIN. $\mathcal{A}$ may also launch denial-of-service attacks to disrupt the pairing process. Nevertheless, repeated failures can trigger alerts from the helper, prompting the user to investigate or report the attack.

\section{Pairing Operations and Evidence}
\label{sec:evidence}

We introduce pairing operations in Section~\ref{subsec:contact}, present universal operation sensing in Section~\ref{subsec:close}, and describe evidence extraction in Section~\ref{subsec:extract_points_events}.

\af
\subsection{Pairing Operations} 
\label{subsec:contact}
To develop efficient pairing procedures, we consider the user interface characteristics of IoT devices. Our survey reveals that the prevalent UI types for resource-constrained IoT devices are buttons (\eg AWS IoT Button~\cite{aws_button}), knobs (\eg Nest Thermostats~\cite{nest}), and touchscreens (typically small; \eg Honeywell T9 Smart Thermostats~\cite{honeywell_thermostat}). Thus, our design of UOS accommodates these UI types and incorporates the following pairing operations.

\begin{itemize}
    \item \emph{Pressing the button a few times}. 
    A \emph{``pause''} here denotes an intentional hold after pressing the button before releasing it, distinct from the natural pause during regular button pressings. Our experiments demonstrate that UOS with pauses \emph{enhances resilience to trained mimicry attacks} compared to UOS without pauses (Section~\ref{subsec:security}).
    
    \item \emph{Twisting the knob back and forth}. When the knob is twisted, the microcontroller on the IoT device can detect both the direction and extent of the twisting motion. To introduce a pause, the user intentionally holds for a random short time right before changing the twisting direction.

    \item \emph{Zig-zag swiping on the touchscreen}. Instead of requiring the user to draw a specific shape or pattern on a small screen, which can negatively impact usability, the user can simply swipe the screen with a finger from left to right and back again several times. Similarly, to enhance security, the user can briefly hold the finger in place right before changing the swiping direction.
\end{itemize}
\af

All the operations are straightforward and user-friendly. Importantly, each operation involves \emph{``crispy''} speed/direction changes, which can be detected by both the IoT device and the helper (Section~\ref{subsec:close}).

\begin{figure*}[ht]
\centering
\includegraphics[scale=0.28]{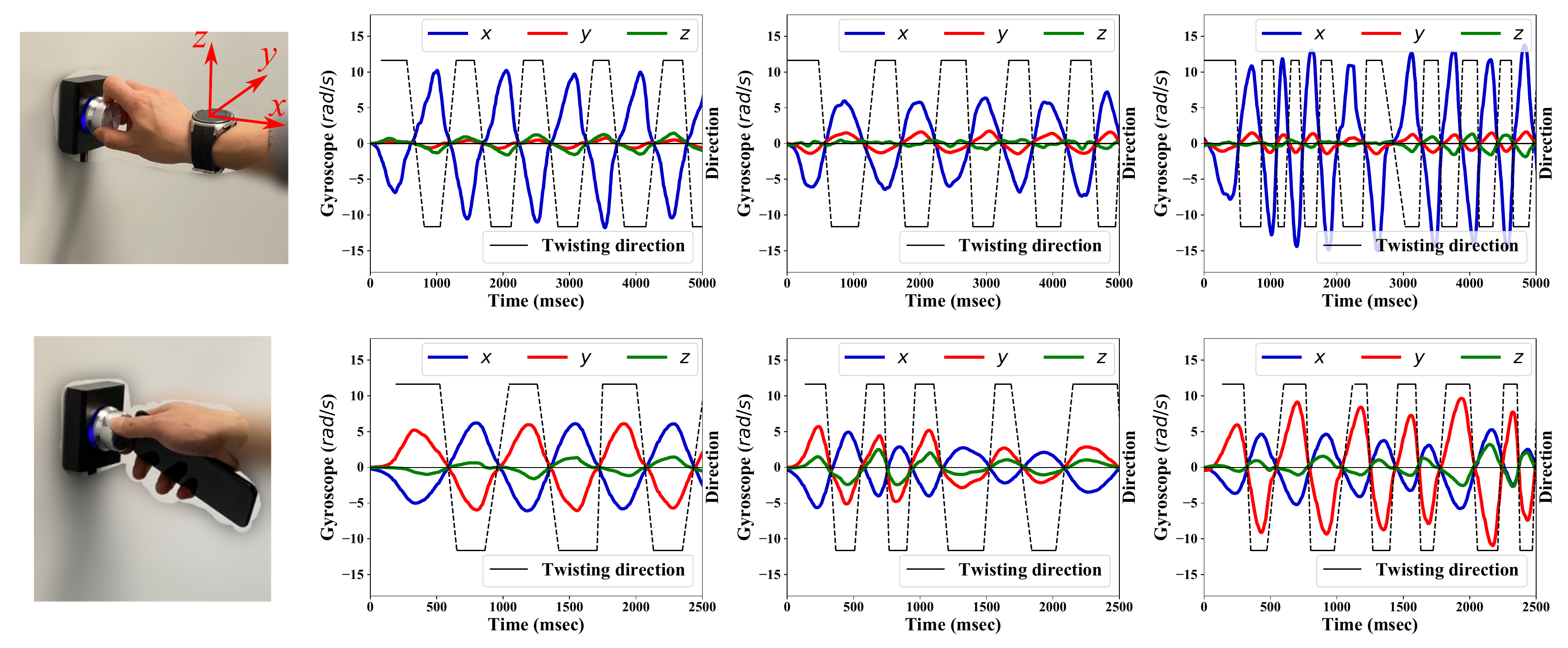} \\
\begin{tabular}{P{3.5cm} P{4.0cm} P{4.4cm} P{4.5cm}}
\footnotesize
Demonstration & \footnotesize User 1 & \footnotesize User 2 & \footnotesize User 3
\end{tabular}
\aaf\aaf
\caption{Gyroscope data captured when three users twist knobs. 
The black lines show
the ground truth of twisting direction.} 
\label{fig:motion_data_knob}
\aaf
\end{figure*}

\af
\subsection{Universal Operation Sensing} \label{subsec:close}
On the IoT device side, it is straightforward to use its controller or sensors to detect button presses, knob twists, or screen swipes.
On the helper side, motion data is collected using the embedded IMU during pairing operations. This prompts exploration into the following questions.
(1) Does the IMU data exhibit correlations with the ground truth?
(2) Are these correlations consistent across different devices, users, and pairing instances? 

To this end, participants are requested to perform each of the three pairing operations. They have the flexibility to choose their hand and wrist posture and can utilize different helpers, such as a smartwatch or smartphone. For instance, the image in the upper left corner of Figure~\ref{fig:motion_data_knob} depicts a user wearing a smartwatch, while the one in the lower left corner shows a user holding a smartphone. 

In the context of knob twisting, as demonstrated in Figure~\ref{fig:motion_data_knob}, a strong correlation is observed between twisting operations and gyroscope data, irrespective of the user or their hand/wrist posture. The gyroscope data exhibits a transition from positive to negative values (and vice versa) in angular velocity as the rotation direction changes (black line), reflecting the ground truth. This correlation is evident in at least one axis of the gyroscope data, illustrated by the $x$-axis (blue line) in the upper row of Figure~\ref{fig:motion_data_knob}, and both the $x$-axis (blue line) and $y$-axis (red line) in the lower row. Identifying the axis with the most significant value changes is straightforward, and we refer to it as the \emph{dominant axis}.

Similar strong correlations are observed for button pressing and zig-zag swiping operations (see figures
in Appendix, available in the online supplemental material). 
When the user's finger presses a button, the acceleration quickly peaks as the finger's movement speed abruptly drops to zero, while the gyroscope data remains largely unaffected by this action. Like twisting knobs, during zig-zag swiping, gyroscope data shows significant changes corresponding to direction changes.

\begin{figure*}[!htb]
\centering
\aaf
\subfloat[\footnotesize{Pairing via pressing a button}]{\includegraphics[scale=0.382]{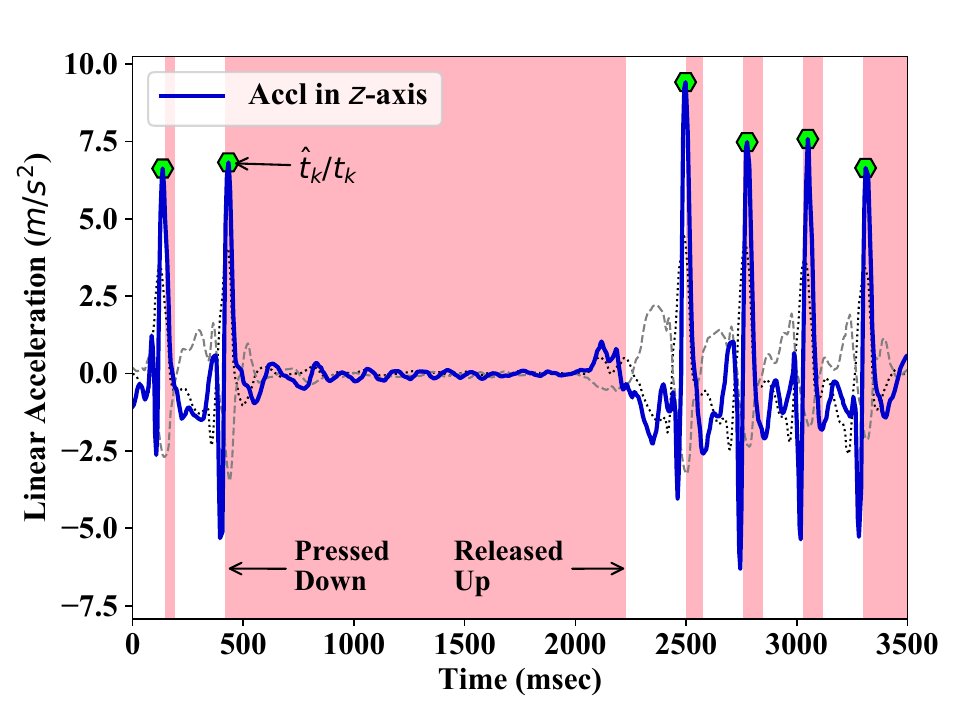}}
\subfloat[\footnotesize{Pairing via twisting a knob}]{\includegraphics[scale=0.358]{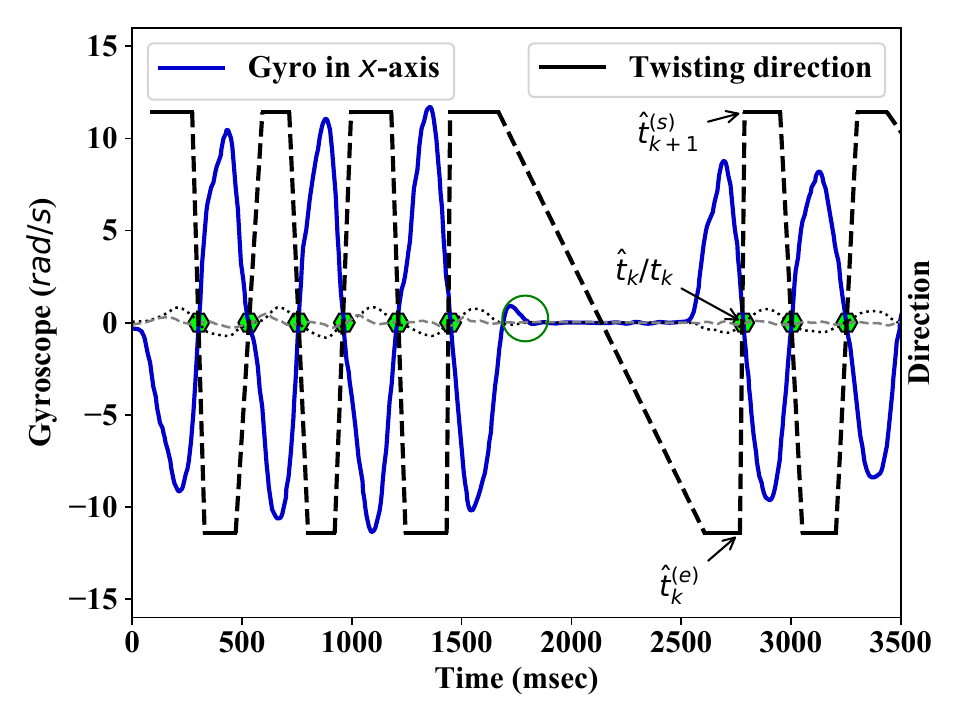}}
\subfloat[\footnotesize{Pairing via swiping a touchscreen}]{\includegraphics[scale=0.358]{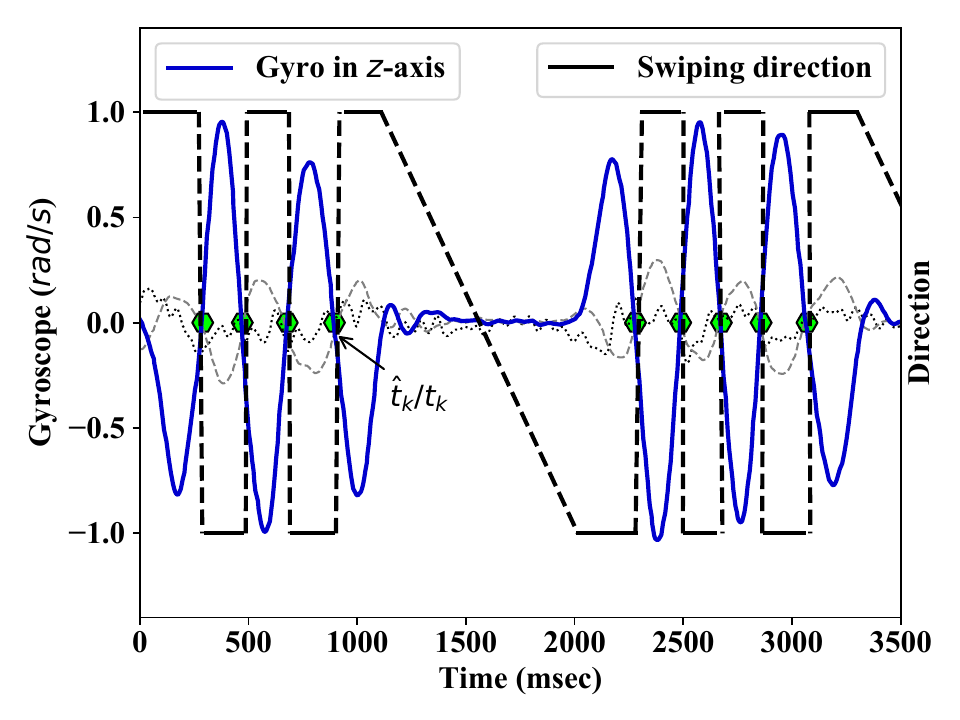}}~~~

\caption{Salient points for the three types of pairing operations. A pause is involved in each type of pairing operations shown in the subfigure.}
\label{fig:gesture_data_sample}
\aaf\af
\end{figure*}

\af\vspace{-1pt}
\subsection{Extracting Evidence}
\label{subsec:extract_points_events}
While strong correlations exist, directly comparing the two heterogeneous data streams is challenging: the IoT device receives input events, whereas the helper's IMU produces motion data. To address this, we extract \emph{salient points} from both and use their \emph{occurrence times} to represent them. We denote the IoT device as $d_1$ and the helper as $d_2$.

\subsubsection{Salient Points on the IoT Device Side}
\label{subsubsec:events_on_iot}

\noindent \textbf{\newline Pressing buttons.} 
Pressing a button once triggers two events: \texttt{PressedDown} and \texttt{ReleasedUp}, as depicted in Figure~\ref{fig:gesture_data_sample}(a). The duration between two consecutive \texttt{PressedDown} and \texttt{ReleasedUp} events is highlighted in pink. During pairing, we utilize the \texttt{PressedDown} events as the \emph{salient points} since they are detectable on both ends (refer to Section~\ref{subsec:close}). Consequently, we derive the timestamp sequence $S_{d_1} = \{\hat{t}_1, \hat{t}_2, \dots , \hat{t}_n\}$, where $\hat{t}_k$ denotes the occurrence time of the $k$th \texttt{PressedDown} event. It is important to note that a random pause only extends the time span between two consecutive salient points, and thus, we do not explicitly identify and represent pauses.

\vspace{2pt}
\noindent \textbf{Twisting knobs.} 
Each rotation-direction change is considered a salient point, as depicted in Figure~\ref{fig:gesture_data_sample}(b). The $k$th salient point is represented by $\hat{t}_{k} \approx \frac{1}{2}(\hat{t}_k^{(e)}+\hat{t}_{k+1}^{(s)})$, where $\hat{t}_k^{(e)}$ denotes the end time of the $k$th rotation, and $\hat{t}_{k+1}^{(s)}$ denotes the start time of the $(k+1)$th rotation. The timestamps $\hat{t}_k^{(e)}$ and $\hat{t}_{k+1}^{(s)}$ should be close for identifying a salient point. Thus, we obtain $S_{d_1} = \{\hat{t}_{1}, \hat{t}_{2}, \dots , \hat{t}_{n-1} \}$, where $\hat{t}_k$ represents the occurrence time of the $k$th salient point.

\vspace{2pt}
\noindent \textbf{Swiping touchscreens.} 
Each swiping-direction change is considered a salient point, depicted in Figure~\ref{fig:gesture_data_sample}(c). We extract a timestamp sequence $S_{d_1} = \{\hat{t}_{1}, \hat{t}_{2}, \dots , \hat{t}_{n-1} \}$, where $\hat{t}_{k}$ represents the $k$th salient point.

\subsubsection{Salient Points on the Helper Side}
\label{subsubsec:events_on_wristband}

\noindent \textbf{\newline Pressing buttons.}
Figure \ref{fig:gesture_data_sample}(a) illustrates the pairing process via pressing a button. In this scenario, the $z$-axis of acceleration serves as the dominant axis (see Section~\ref{subsec:close}), while the signals along the other two axes are depicted in dashed grey lines. At each salient point of the ground truth, marked by a \texttt{PressedDown} event, a sharp peak is evident. We extract the occurrence time of each sharp peak and derive the sequence $S_{d_2} = \{t_1, t_2, \dots , t_m \}$, where $t_k$ represents the time of the $k$th salient point.

\vspace{1pt}
\noindent \textbf{Twisting knobs.}
Based on our investigation of motion data (refer to Section~\ref{subsec:close}), we utilize gyroscope data to detect salient points, which signify rotation-direction changes. In the example depicted in Figure~\ref{fig:gesture_data_sample}(b), the $x$-axis emerges as the dominant axis. As the rotation direction of the IoT device shifts, the gyroscope signal undergoes a corresponding change in sign. Hence, we identify salient points by pinpointing instances of signal sign transitions during substantial amplitude alterations. Subsequently, we extract a sequence of timestamps for all detected salient points, denoted as $S_{d_2} = \{t_1, t_2, \dots , t_m \}$, where $t_k$ denotes the occurrence time of the $k$th salient point.

During a pause, the gyroscope readings typically hover around zero. However, slight fluctuations may still occur, particularly at the outset (as indicated by the large green circle in Figure~\ref{fig:gesture_data_sample}(b)). To prevent the identification of false salient points, we apply a straightforward thresholding method to filter out such fluctuations.

\vspace{1pt}
\noindent \textbf{Swiping touchscreens.}
Every change in swiping direction generates a salient point in the gyroscope data trace. Illustrated in Figure \ref{fig:gesture_data_sample}(c), each salient point aligns with a sharp sign change due to a shift in swiping direction. Consequently, we derive a sequence of timestamps: $S_{d_2} = \{t_1, t_2, \dots , t_m \}$, where $t_k$ represents the time of occurrence for the $k$th salient point.

\vspace{2pt}
\noindent \textbf{Big silence.}
Identifying the first salient point is crucial. Following the initiation of pairing, such as through a long press on a button, the motion data may exhibit noisy motion data resembling salient points as the user's hand approaches the IoT device's button, knob, or screen. To mitigate this issue, we instruct the user to hold their hand on the button, knob, or screen for approximately 2 to 3 seconds before performing the pairing operations. This deliberate \emph{``big silence''} in the motion data serves as a clear indication that pairing operations will follow, allowing the detection of salient points from the motion data to commence.

\subsubsection{No Clock Synchronization}
\label{subsec:rep_evidence}
To eliminate the need for clock synchronization, we convert each timestamp sequence into a series of time intervals using the equations $\hat{i}_k = \hat{t}_{k+1} - \hat{t}_{k}$ and $i_k = t_{k+1} - t_{k}$ for $S_{d_1}$ and $S_{d_2}$, respectively. We then concatenate the time intervals and refer to them as \emph{evidence}: $E_{d_1} = \{\hat{i}_1||\hat{i}_2||\cdots||\hat{i}_{q-1} \}$ and $E{d_2} = \{i_1||i_2||\cdots||i_{p-1} \}$, where $E_{d_1}$ represents the evidence collected by the IoT device, and $E_{d_2}$ by the helper. Note that clock drift during pairing is not a concern, as the pairing process lasts only about three seconds, resulting in a maximum drift of approximately three milliseconds~(see Section~3.2 of~\cite{mani2018system}), which is well within the tolerance of our protocol.

\section{\shortname: Pairing with Information Loss}
\label{sec:protocol}
Once the two pieces of evidence are extracted, both sides use them to establish a shared key.
This section presents a fuzzy commitment-based pairing protocol, implemented in a system named \shortname. 

\af
\subsection{Faithful Fuzzy Commitment} 
\label{subsec:challenges_and_solutions}
\noindent\textbf{Challenges.} Performing secure mutual evidence verification in the presence of powerful attacks like MITM attacks poses a significant challenge. Additionally, small differences between the observations of salient points from the wristband and the IoT device, such as those arising from sensor readings and clock drift, present another challenge.

\vspace{2pt}
\noindent\textbf{Failed attempt.} 
To tackle these challenges, we initially employ a fuzzy commitment scheme based on error-correcting codes~\cite{Juels_FCS_CCS99}. The fuzzy commitment scheme, previously used in proximity-based pairing~\cite{audio_pairing, Perceptio, 2014CCS_CZP}, enables mutual evidence verification without revealing the evidence to MITM attackers and accommodate small differences between two pieces of evidence. In this scheme, the sender encrypts its evidence into a message that can be successfully decrypted only if the receiver possesses evidence that is sufficiently similar to the sender's evidence based on the metric of Hamming distance~\cite{Juels_FCS_CCS99}. 
We refer to the original fuzzy commitment as the \emph{vanilla} fuzzy commitment.

To conduct fuzzy commitment, the evidence must first be encoded into a bit representation. 
Previous studies convert values directly to their binary representations~\cite{Perceptio}. However, this encoding method may incorrectly categorize two dissimilar (resp.\ similar) evidence sequences as similar (resp.\ dissimilar).

For instance, considering the interval values \{121\} and \{57\}, encoded as ``0111 1001'' and ``0011 1001'' respectively, the Hamming distance, which is the count of different digits in the two bit strings, is $\mathrm{Ham}(121, 57) = 1$. Due to their small Hamming distance, the two intervals are deemed \emph{similar}, whereas their actual difference is significant.
Similarly, for the interval values \{128\} and \{127\}, represented as ``1000 0000'' and ``0111 1111'' respectively, the Hamming distance is $\mathrm{Ham}(128, 127) = 8$. Consequently, the algorithm incorrectly categorizes these two similar interval values as very \emph{different}.

\begin{table*}[!ht]
    \centering
 \caption{Faithful fuzzy-commitment-based pairing protocol, which causes information loss and is used in \shortname{}.}
    \aaf \af
    \setlength{\extrarowheight}{4pt}
    \setlength{\tabcolsep}{5pt}
    \renewcommand{\arraystretch}{1}
    \scalebox{0.85}{
    \begin{tabular}{|p{7cm}P{2.8cm}p{7cm}|}
    \hline
    IoT device $d_1$    &   &   Helper device $d_2$  \\ \hline
    \multicolumn{3}{|c|}{Phase 1: Initialization}  \\ 
    & \emph{Initiates the pairing} & \\ \hline
    
    \multicolumn{3}{|c|}{Phase 2: Extracting evidence}  \\ 
    extracts evidence $E_{d_1}$   &   &   extracts evidence $E_{d_2}$  \\ 
    if self-checking fails, aborts  &  & if self-checking fails, aborts and reminds the user\\ \hline
    
    \multicolumn{3}{|c|}{Phase 3: Fuzzy Commitment}  \\
   \circled{1}  picks a random value $P \in \mathbb{F}_{2^k}^m$  &   &  \\
   \circled{2}  $\lambda \in \mathbb{F}_{2^k}^n$  $\xleftarrow[]{\text{encode}}$ $\mathrm{RS}(2^k, m, n, P)$  &   &   \\
   \circled{3} commits: $\delta = e(E_{d_1}) \oplus \lambda$  &   $\xrightarrow{\quad \delta \quad}$   &    \circled{4} decommits: $\lambda' = e(E_{d_2}) \oplus \delta$ \\
   &   &  \circled{5} $P'$ $\xleftarrow[]{\text{decode}}$ $\overline{\mathrm{RS}}(2^k, m, n, \lambda')$  \\ \hline
    \multicolumn{3}{|c|}{Phase 4: PAKE}  \\
    \circled{6} picks $a$; $A = g^a \bmod p$; $w=\mathrm{h}(P)$  & $\xrightarrow{\quad \mathrm{E}(w,\, A) \quad}$ & \circled{7} picks $b$; $B=g^b \bmod p$; $w'=\mathrm{h}(P')$ \\
    
    \circled{9}  $K = B^a \bmod p$ & $\xleftarrow{ \mathrm{E}(w',\, B||C_1) }$ & \circled{8} $K'=A^b \bmod p$; picks a challenge $C_1$\\
    
    \circled{10} picks a challenge $C_2$ &  $\xrightarrow{\mathrm{E}(K,\, C_1||C_2) }$ & \circled{11} if $C_1$ is not received, aborts\\
    \circled{12} if $C_2$ is not received, aborts    &  $\xleftarrow{\; \mathrm{E}(K',\, C_2) \;}$ & \\
    \hline
    \end{tabular}
    }
    \label{tab:protocol_chart}
    \aaf
\end{table*}

\vspace{2pt}
\noindent\textbf{Solution.} 
To tackle this issue, we propose \emph{faithful fuzzy commitment}. It encodes each time interval by dividing the interval value by a base value, thereby accommodating small differences and reducing the encoding length. The result is represented as a sequence of \emph{consecutive} ``1'' and ``0'' bits. The Hamming distance between these encodings can then be computed to determine similarity.

Given the base value $B$  and an interval value $i$, we compute $n=\left \lfloor{i/B}\right \rfloor$. 
Ensuring all intervals have the same encoding length $L$, the interval is represented as $n$ \emph{consecutive} ``1'' bits, with an additional $L-n$ ``0'' bits appended to the end. Thus, the encoding for an interval value $i$ is:

\aaf
\begin{equation}
    e(i) = \underbrace{\overbrace{1, 1, \cdots, 1}^{n}, 0, 0, \cdots, 0}_{L}
\end{equation}
\aaf

A larger base value $B$ results in more efficient key agreement but less precise evidence comparison, and vice versa. 
The selection of the base value $B$ is discussed in Section~\ref{subsec:stability}.
For instance, with $B=4$, consider the two examples above. The interval \{121\} can be encoded as $\left \lfloor{121/4}\right \rfloor=30$ consecutive ``1'' bits followed by $L-30$ ``0'' bits. Similarly, \{57\} can be encoded as 14 consecutive ``1'' bits followed by $L-14$ ``0'' bits. Consequently, we have $\mathrm{Ham}(e(121),~e(57)) = 16$. Likewise, based on our encoding algorithm, 
$\mathrm{Ham}(e(128),~e(127)) = 1$.

\vspace{2pt}
\subsection{Pairing Protocol} \label{subsec:details}

Table~\ref{tab:protocol_chart} presents our protocol, which consists of four phases:  
(1) \emph{Initialization}. Most commercial off-the-shelf devices include a built-in method to initiate pairing, such as long-pressing a button.  
(2) \emph{Extracting Evidence}. As the user performs pairing operations on the IoT device while wearing or holding the helper device, both sides independently extract evidence. During this phase, \emph{self-checking} is enforced: if no pauses are detected, the pairing aborts and the helper prompts the user to introduce one or more pauses.  
(3) \emph{Fuzzy Commitment}. The two devices use the evidence to transmit a ``password''.  
(4) \emph{Password-Authenticated Key Exchange} (PAKE). The devices use the ``password'' to derive a shared key.  
Below, we elaborate on Phases (3) and (4).

\zm{\noindent \textbf{Fuzzy Commitment}. We employ a faithful fuzzy commitment scheme based on Reed-Solomon (RS) codes~\cite{rs_codes, Juels_FCS_CCS99}. Device $d_1$ selects a random value $P$, encodes it into a codeword $\lambda$ using RS encoding, and computes a commitment $\delta = e(E_{d_1}) \oplus \lambda$, where $E_{d_1}$ is the evidence and $e()$ is a feature extractor. Device $d_2$ uses its own evidence $E_{d_2}$ to recover $\lambda' = e(E_{d_2}) \oplus \delta$ and attempts to decode $P'$. If $E_{d_1}$ and $E_{d_2}$ are similar, $P' = P$; otherwise, decoding fails.

Since $P$ may be vulnerable to brute-force guessing (BF-offline), it cannot serve directly as a shared key. Instead, we treat $P$ as a password input to a PAKE protocol~\cite{pake-proof} to derive a secure session key.}

\vspace{2pt}
\zm{{\noindent \textbf{PAKE.} We adopt the Diffie-Hellman Encrypted Key Exchange (DH-EKE) protocol~\cite{pake}, a member of the PAKE family standardized in IEEE P1363.2~\cite{1363.2}, though other PAKE variants could also be used. DH-EKE prevents MITM attacks by encrypting both $A$ and $B$ using a shared password when sending messages in \circled{6} and \circled{8}. The base $g$ and modulus $p$ are public parameters; $\mathrm{h}()$ is a cryptographic hash, and $\mathrm{E}()$ denotes symmetric encryption. If $P' \neq P$, $d_2$ receives a value different from its challenge $C_1$ (\circled{11}); otherwise, both $d_1$ and $d_2$ derive the same session key $K = K'$ via steps \circled{11} and \circled{12}.}}

\vspace{2pt}
\zm{\noindent \textbf{Parameter Consideration.}
The security of $\lambda$ primarily depends on the size of the codeword space ($2^k$)~\cite{Juels_FCS_CCS99}; for robust security, $k$ should exceed 80. Using RS encoding, a word of length $m$ is uniquely mapped to a codeword of length $n$, allowing correction of up to $Thr = \lfloor{\frac{n - m}{2}}\rfloor$ bits. The key $P'$ can be correctly recovered only if $\mathrm{Ham}(e(E_{d_1}), e(E_{d_2})) \leq Thr$. We discuss the choice of $Thr$ in Section~\ref{subsec:performance}.}

\vspace{2pt}
\noindent \textbf{Resilience to Attacks.}
The forward secrecy property of DH ensures that even if $P$ is cracked offline (e.g., by analyzing recorded video of the user or exhaustively trying all possible evidence), it cannot be used to reconstruct the session key. As a result, offline brute-force attacks are ineffective. Online attacks are also futile: PAKE provides zero-knowledge password proofs~\cite{1363.2}, allowing an active MITM attacker only a single guess, regardless of their computational power, and revealing no information unless the guess is correct.

\vspace{2pt}
\noindent \zm{\textbf{Summary and Limitations.} 
The protocol uses faithful fuzzy commitment to mitigate inaccuracy and PAKE to derive a high-entropy key. However, it has the following limitations:  
(1) Interval value encoding still results in information loss and, hence, inaccuracy. For example, with $B = 10$, the values 120 and 129, though quite different, are encoded identically, while 128 and 130, despite being similar, produce different encodings.  
(2) Selecting an appropriate base value $B$ involves a trade-off: a small $B$ increases sensitivity to small differences and produces longer encodings, leading to higher false rejection rates and computation overhead; a large $B$ leads to higher information loss and greater susceptibility to mimicry attacks.}
\section{\precise: Pairing with Zero Information Loss}
\label{sec:timing_protocol}

\begin{table*}[!ht]
    \centering
    \caption{\pname pairing protocol, which avoids information loss and is used in \precise.}
    \aaf \af
    \setlength{\extrarowheight}{4pt}
    \setlength{\tabcolsep}{5pt}
    \renewcommand{\arraystretch}{1}
    \scalebox{0.85}{
    \begin{tabular}{|p{8.5cm}P{2.8cm}p{8.5cm}|}
    \hline
    IoT device $d_1$    &   &   Helper device $d_2$  \\ \hline
    \multicolumn{3}{|c|}{Phase 1: Initialization} \\
        & \emph{Initiates the pairing} & \\ 
    \multicolumn{3}{|c|}{\emph{Diffie-Hellman key exchange}} \\
    picks $a$; $A = g^a \bmod p$ & $\xrightarrow{A}$  & \\
    & $\xleftarrow{B}$  & picks $b$; $B = g^b \bmod p$ \\
    $K = B^a \bmod p$ & & $K = A^b \bmod p$
    \\ \hline
    \multicolumn{3}{|c|}{Phase 2: Extracting evidence}  \\
    extracts evidence $E_{d_1}$   &   &   extracts evidence $E_{d_2}$  \\ 
    if self-checking fails, aborts  &  & if self-checking fails, aborts and reminds the user\\ \hline
    
    \multicolumn{3}{|c|}{Phase 3: Mutual authentication}  \\
    \multicolumn{3}{|c|}{(all the messages below are encrypted with $K$)} \\
    
    $t_1$: the end time of the pairing operations  &  & $t_2$: the end time of the pairing operations \\

    \circled{1}  picks $w_1$; $C_1 \xleftarrow{\quad} \texttt{Commit}((K || E_{d_1}); w_1)$  & $\xrightarrow{C_1}$ & \\ 
    & $\xleftarrow{C_2}$ & \circled{2} picks $w_2$; $C_2 \xleftarrow{\quad} \texttt{Commit}((K || E_{d_2}); w_2) $\\ 
 & & \circled{3} aborts if  $C_1$ is not received by the deadline $t_2 + T_{thr}$  \\
    \circled{4} aborts if $C_1 == C_2$  &  & \\
    waits until after the deadline $t_1 + T_{thr}$ & $\xrightarrow{w_1}$ & \circled{5} gets $(E_{d_1}^\prime, K^\prime)$ from $C_1$ \\
    & $\xleftarrow{w_2}$  & aborts if ($K^\prime \neq K$) or ($E_{d_1}^\prime$ and $E_{d_2}$ do not correlate) \\
    \circled{6} gets $(E_{d_2}^\prime, K^\prime)$ from $C_2$& & \\ 
    aborts if ($K^\prime \neq K$) or ($E_{d_1}$ and $E_{d_2}^\prime$ do not correlate) &   &  \\ 
    \hline
    \end{tabular}
    }
    \label{tab:timing_protocol_chart}
    \aaf
\end{table*}

\zm{This section presents the \pname pairing protocol that eliminates the limitations aforementioned.  The pairing system running 
this protocol is named \precise.}

\vspace{2pt}
\noindent \textbf{Main Idea.} A MITM attack typically works by \emph{first} obtaining the secret (evidence, in our case) from one side (e.g., the IoT device), and \emph{then} using it to generate a valid commitment to fool the other side. Once the secret is known, the attacker can create a correct commitment.  

To defeat such attacks, we propose the notion of a \emph{commitment deadline}: a commitment must be received \emph{before} a specified deadline to be accepted, while the decommitment must be conducted \emph{after} the deadline. This enforces two properties: (1) the attacker cannot obtain the pairing information before the deadline, and (2) any commitment received after the deadline, regardless of correctness, is discarded.  
This protocol adopts a philosophy fundamentally different from fuzzy commitment. Rather than concealing the secret using special encoding that causes information loss, we rely on standard encryption or hash for commitment.


\vspace{2pt}
\noindent \textbf{Protocol Details.}
The overall design of the \pname pairing protocol is to first establish a key-protected communication channel, 
which is then authenticated using the secrets on the two sides.
Table~\ref{tab:timing_protocol_chart} shows the pairing protocol, which contains three phases. (1) \emph{Initialization}. After the pairing process is initiated (\eg by long pressing a button), the two device negotiates a key $K$ through Diffie-Hellman key exchange. 
As the device(s) may be fooled to communicate with an attacker, the communication channel needs to be authenticated. 
All the subsequent messages are encrypted and authenticated using the key.
(2) \emph{Extracting evidence}. This phase is the same as that in Table~\ref{tab:protocol_chart}. 
(3) \emph{Mutual authentication}. The two devices exchange their evidence to make an authentication decision, which is elaborated below.

Specifically, after the physical operations on the devices are completed, both devices record the timestamps, denoted as $t_{1}$ and $t_{2}$. $d_{1}$ randomly selects a key $w_{1}$, computes
the commitment $C_1 = \texttt{Enc}_{w_1}(K || E_{d_1})$ and sends $C_{1}$ to $d_{2}$ \emph{immediately} ($\circled{1}$).\footnote{An alternative design is to
send a cryptographic hash of $K || E_{d_1}$ as the commitment, and the decommitment reveals $K || E_{d_1}$.} Meanwhile, $d_{2}$ also randomly selects a key $w_{2}$, computes
the commitment $C_2$ and sends it to $d_1$ ($\circled{2}$). 
The helper aborts if $C_1$ is not received by the deadline $t_2 + T_{thr}$, where $T_{thr}$ is a \emph{threshold} discussed below ($\circled{3}$). After receiving $C_2$, 
if $C_2$ is equal to $C_1$,
$d_1$ aborts the process, which is to prevent attacks that simply send $C_1$ back to fool $d_1$.
Otherwise, $d_1$ \textbf{waits until after the deadline} $t_1 + T_{thr}$ to send the key $w_1$ ($\circled{4}$). In short, $d_1$ does not decommit  until $C_2$ is received \emph{and} the deadline has expired.

After receiving $w_1$, $d_2$ decrypts $C_1$ to get $E_{d_1}^\prime$ and  $K^\prime$.
The process aborts if $K^\prime \neq K$ or $E_{d_1}^\prime$ and $E_{d_2}$ do not correlate. Otherwise, 
$d_2$ sends the key $w_2$ to $d_1$ ($\circled{5}$). $d_1$ then performs the same checking ($\circled{6}$). A success means that
the communication channel with the key $K$ has been authenticated by both parties and thus
they can continue using this channel for communication. 

\zm{The protocol requires a fixed threshold $T_{thr}$, determined by the time needed for $d_1$ to generate (we use AES-128 encryption) and transmit the commitment. To ensure broad applicability and minimize false rejections even for low-end devices and low-speed data links, we set this threshold conservatively. For example, in the MSP432 ultra-low-power microcontroller (commonly used in low-end IoT devices), hardware-accelerated AES takes only 0.5\,ms~\cite{iot_encryption}. In another low-resource platform, the PIC18F27K40 (\~{}\$2), software-based AES requires just 75.6\,ms~\cite{iot_encryption}. 
We therefore select a conservative threshold of $T_{thr} = 600$\,ms. This accommodates delays and jitter typical of IoT communication technologies such as Wi-Fi, Bluetooth, and Zigbee. For instance, real-world Wi-Fi latency is under 55\,ms at distances below 10 meters~\cite{sultan2022real}, while Bluetooth and Zigbee typically introduce delays below 100\,ms~\cite{blt_zigbee}. BLE often achieves latency under 10\,ms~\cite{tosi2017performance}. Therefore, the 600 ms threshold provides a robust margin  for reliable operation without introducing noticeable latency to the user experience.

Since the helper and IoT device remain in close proximity during pairing and communicate over a short-distance local network, jitter and packet loss are minimal~\cite{sheshadri2017packet}.
In rare cases, due to excessive delays or packet loss, false rejections can arise; in such instances, our system aborts pairing and prompts users to re-initiate the process. 
An active attacker who introduces excessive delays, jitter, and packet loss effectively mounts a denial-of-service attack. As noted in our Threat Model (Section~\ref{sec:system_overview}), repeated failures can trigger alerts from the helper.}

\crm{It is worth noting that the protocol is \emph{asymmetric}, in that the commitment deadline is enforced on the IoT device side rather than on the helper device side. Specifically, the IoT device $d_1$ must send its commitment within $T_{thr}$, whereas the helper $d_2$ may do so after $T_{thr}$. (To prevent DoS exploitation, a relatively large threshold, e.g., five seconds, can be applied to $d_2$.) This design stems from the observation that the time required for commitment generation and transmission on IoT devices tends to be stable and consistent, while helper devices such as smartwatches or smartphones often exhibit greater timing variability due to their multitasking environments. This asymmetric design makes the protocol
both efficient and widely applicable.}


\vspace{2pt}
\noindent \textbf{Correlation Checking.}
The correlation of the evidences is determined by a machine learning classifier. Specifically, given a pair of evidence $E_{d_{1}} = {i_{1}^{1}, i_{2}^{1}, ..., i_{l}^{1}}$ and $E_{d_{2}} = {i_{1}^{2}, i_{2}^{2}, ..., i_{l}^{2}}$, where $l$ denotes the evidence length, we first calculate the absolute difference of each element. Based on the absolute difference vector, we derive the following features inspired by P2Auth~\cite{p2auth}: (1) \emph{Average}: the average of the elements; (2) \emph{Standard deviation}: the standard deviation of the elements; (3) \emph{Minimum}: the minimum of the elements; (4) \emph{Maximum}: the maximum of the elements; (5) \emph{Max-Min difference}: the difference between the maximum and minimum of the elements; (6) \emph{MAD}: median absolute deviation of the elements; (7) \emph{Modified z-score}: the modified z-score of the elements. These features are then formed as a feature vector, which is the input of the machine learning classifier. The evidence length $l$ is studied in Section~\ref{subsec:security}.

\vspace{2pt}
\noindent \textbf{Resilience to Attacks.}
The key $K$ is generated via a Diffie-Hellman (DH) key exchange. Since $K$ is independent of the evidence, even if $E_{d_1}$ or $E_{d_2}$ is obtained after pairing, it cannot be used to reconstruct the session key.

One may consider a MITM attacker $A$ positioned between $d_1$ and $d_2$, attempting to deceive both parties. Let $K_1$ be the key established between $d_1$ and $A$, and $K_2$ the key between $A$ and $d_2$. Critically, the commitment $C_1$ sent from $d_1$ to $A$ is $\texttt{Enc}_{w_1}(K_1 \,\|\, E_{d_1})$ (Step $\circled{1}$). Without the decryption key $w_1$, which is not revealed to $A$ until after the deadline, $A$ cannot extract $E_{d_1}$ from $C_1$, and thus cannot generate a valid $C_1^\prime = \texttt{Enc}_{w_1^\prime}(K_2 \,\|\, E_{d_1})$, where $w_1^\prime$ is a key chosen by $A$, and $C_1^\prime$ is sent to $d_2$ (Step $\circled{1}$).


\begin{table}
    \centering
    \small
    \renewcommand{\arraystretch}{1.0}
    \caption{\crm{Comparison with prior pairing protocols.}} 
    \label{tab:comparison}
        \aaf\af
    \resizebox{0.48\textwidth}{!}{%
    \begin{tabular}{c|c|c}
    \hline
    \textbf{Pairing Protocols}  &  \textbf{Secure?}  &  \textbf{Zero Information Loss?} \\  
    \hline
    Interlock~\cite{rivest1984expose} & \xmark & \cmark \\ \hline
    H2H~\cite{H2H_CCS14} & \xmark & \cmark \\ \hline
    Fuzzy commitment (used in \shortname) & \halfcheck & \xmark \\ \hline
    \pname (used in \precise) & \cmark & \cmark \\ \hline
    \end{tabular}
    }
    \vspace{-10pt}
\end{table}


\vspace{2pt}
\noindent\textbf{Comparison with Prior Pairing Protocols.}
\crm{Many pairing protocols have been proposed; however, as summarized in Table~\ref{tab:comparison}, they are either insecure or suffer from inaccuracy due to information loss. 
Interlock~\cite{rivest1984expose} avoids information loss but remains vulnerable to multiple attacks~\cite{wu2017attack}. 
H2H~\cite{H2H_CCS14} is also vulnerable. Specifically, a reflection attack based on identical commitment can be used to launch a man-in-the-middle (MITM) attack. 
In both protocols, an attacker can impersonate one side, Alice, to steal the pairing information from the other side and subsequently use it to deceive Alice.
Fuzzy commitment suffers from information loss and the resulting false acceptance can be exploited by attackers. 
 In contrast,
the \pname pairing protocol is secure
and causes zero information loss. Our evaluation shows that, compared to prior methods, \precise achieves higher accuracy (Section~\ref{subsec:performance}) and demonstrates greater security (Section~\ref{subsec:security}).} 

\section{Prototype Implementation}
\label{sec:implementation}

\noindent \textbf{Helper.} 
The pairing process provides users with the option to either wear a wristband or hold a smartphone. 
We implement the prototypes on two helpers: (1) an LG W200 smartwatch, and (2) a Google Nexus 5X smartphone.
For the smartwatch, we develop an application running on Android Wear to collect motion data, while for the smartphone, we create a mobile application. Both devices are equipped with  
an inertial measurement unit, which includes a triple-axis accelerometer and a triple-axis gyroscope.

\begin{figure}[t]
\graphicspath{ {./figs/} }
\centering

\includegraphics[width=0.98\linewidth]{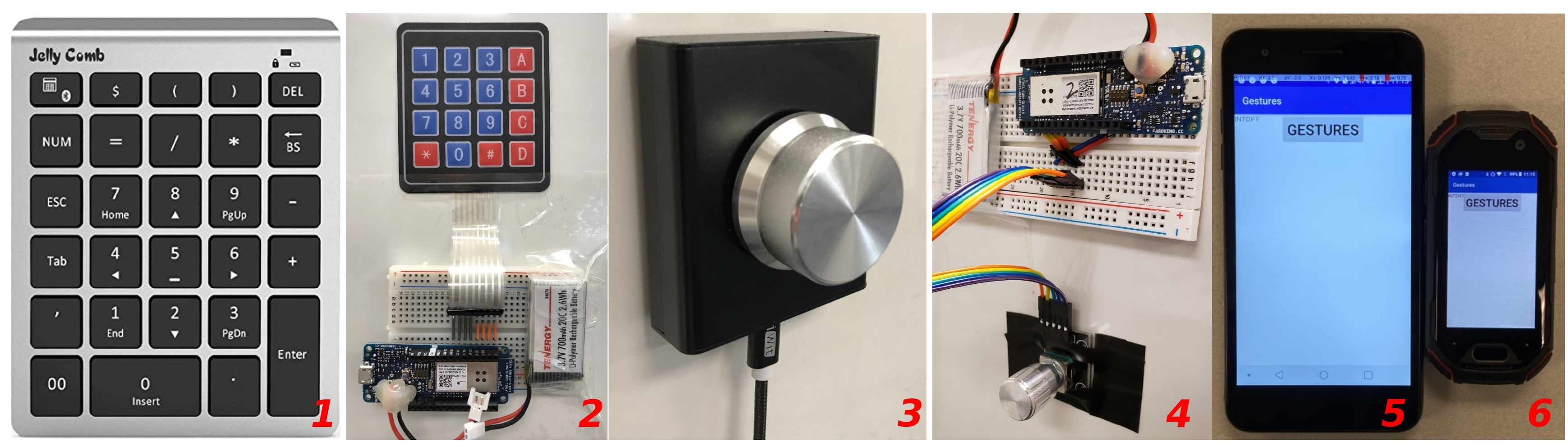} 
\af\aaf
\caption{Six devices are used in our experiments, including two keypads (a plastic keypad labeled as 1, and a rubber one as 2; in either case, we only use \emph{one button} for pairing); 
two knobs (a large knob labeled as 3, and a small one as 4); 
two touchscreens (a 5.2" Google Nexus 5X labeled as 5, and a 2.45" Unihertz Atom labeled as 6).} 
\label{fig:iot_devices}
\aaf\aaf
\end{figure}

\vspace{2pt}
\noindent \textbf{IoT device.} 
A variety of IoT devices are used to build the prototypes, as shown in Fig.~\ref{fig:iot_devices}. 
(1) \emph{Buttons made of two different materials} are used: 
a plastic keypad labeled as 1, and a rubber one labeled as 2. The plastic keypad communicates via a Bluetooth module with the helper, while the rubber one interfaces with an Arduino board MKR1000, communicating through its Wi-Fi module.
(2) \emph{Knobs with two different sizes} are used: a large knob labeled as 3, 
and a small one labeled as 4. The large knob is a volume controller for desktops. An interface function is developed to read data from the large knob, while the small knob is interfaced using an Arduino board MKR1000. 
(3) \emph{Touchscreens with two different sizes} are used: a Nexus 5X, labeled as 5, featuring a screen size of 5.2", and a Unihertz Atom, labeled as 6, with a screen size of 2.45", which has the smallest touchscreen among the smartphones on the market.

\section{Data Collection}
\label{sec:data_collection}
We build two datasets: (1) \emph{Dataset I} measure the accuracy of our system, and (2) \emph{Dataset II} evaluate the resistance of our system against mimicry attacks.

We recruit 20 participants, comprising 14 males and 6 females aged between 18 and 36.  
For data collection, we utilize three devices, including the large knob, the plastic keypad, and the Nexus 5X smartphone. The remaining three devices are reserved for evaluating the system's stability, as discussed in Section~\ref{subsec:stability}.

\af
\subsection{Dataset I for Evaluating Accuracy}
\label{subsec:dataset1}

To construct \emph{Dataset I}, participants are tasked with wearing a smartwatch and performing the pairing operations on each of the three devices, without pauses, for 30 repetitions. Additionally, to evaluate the impact of pauses, participants are instructed to conduct the pairing operations 30 times, introducing either one or two pauses during the process.

\vspace{1pt}
\noindent \textbf{\emph{Positive pairs.}}
When a participant performs the pairing operations on a device, we collect one positive data pair from the smartwatch and device.
Consequently, for pairing operations conducted without pauses, our dataset comprises 1,800 ($ = 20\times 30 \times 3$) positive pairs, each labeled $s=1$. Similarly, for pairing operations with random pauses, we also collect 1,800 ($ = 20\times 30 \times 3$) positive pairs, each labeled $s=1$.

\vspace{1pt}
\noindent \textbf{\emph{Negative pairs.}}
Assume two users, $\mu_1$ and $\mu_2$, perform the pairing operations on two identical devices. 
In this scenario, the evidence $E_{d_1}$ from $\mu_1$'s IoT device and the evidence $E_{h_2}$ from $\mu_2$'s helper constitute a negative pair. Similarly, the evidence $E_{h_1}$ from $\mu_1$'s helper and the evidence $E_{d_2}$ from $\mu_2$'s device form another negative pair.

By randomly selecting two users performing the same pairing operations, 
we generate 1,800 negative pairs (the same amount as the positive pairs) for the pairing operations without pauses, and 1,800 negative pairs for the pairing operations with pauses, each labeled as $s=-1$. 

\subsection{Dataset II for Evaluating Resilience to Mimicry Attacks}
\label{subsec:dataset2}
To build Dataset II, we designate 10 participants as victims and the remaining 10 as attackers. We consider the three attack settings of mimicry attacks as discussed in \textbf{Threat Model} in Section~\ref{sec:system_overview}.

For \textbf{MA-trained}, we initially instruct each victim to perform pairing on each type of device five times while recording a video of each pairing session. Each attacker then undergoes training by watching the corresponding video as many times as needed.
The attacker focuses solely on learning the actions of one specific victim and subsequently launches attacks against that victim. Throughout the training process, we offer immediate feedback to the attackers regarding the disparities between their evidence and that of the victims, enabling them to adjust their operations to mimic more effectively. 

For each attack setting, every pair of attacker and victim conducts the pairing operations, with or without pauses, on each device 15 times. Four pieces of evidence are collected each time: $E_{d_V}$ from the victim's device, $E_{h_V}$ from the victim's helper, $E_{d_A}$ from the attacker's device, and $E_{h_A}$ from the attacker's helper. These pieces are then used to construct two types of evidence pairs based on the attackers' objectives. (\emph{G1}) The first pair consists of $E_{h_V}$ and $E_{d_A}$, indicating that $\mathcal{A}$ attempts to have $\mathcal{V}$'s helper accept a pairing with $\mathcal{A}$'s device. (\emph{G2}) The second pair consists of $E_{d_V}$ and $E_{h_A}$, 
implying that $\mathcal{A}$ attempts to deceive $\mathcal{V}$'s device into pairing with $\mathcal{A}$'s helper.

For each attack setting, 
we collect 900 evidence pairs without pauses,
containing 450 ($ = 10\times 15 \times 3$) \emph{G1} pairs and 450 \emph{G2} pairs.   
The same quantity of pairs is collected for the pairing operations with pauses.

\section{Evaluation}
\label{sec:evaluation}
We conducted four in-lab studies to evaluate our system in terms of pairing accuracy, security, stability, and efficiency. The first study (Section~\ref{subsec:performance}) examines its pairing accuracy, while the second (Section~\ref{subsec:security}) evaluates the resilience of our system to mimicry attacks. The third (Section~\ref{subsec:randomness_entropy}) assesses the randomness and entropy of evidence. The fourth (Section~\ref{subsec:stability}) tests the stability of our system under various parameters and experimental settings. Time efficiency is evaluated in Section~\ref{subsec:efficiency}.

\begin{figure*}[ht]
\aaf\aaf
\centering
\subfloat[Button-based pairing.]{\includegraphics[scale=0.3]{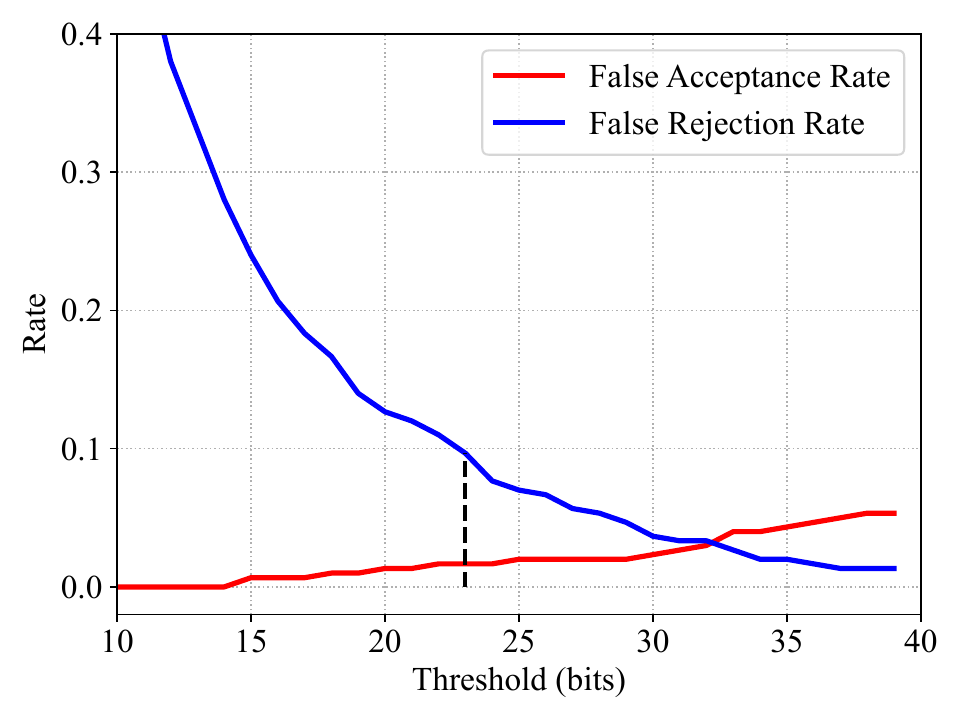}}~
\subfloat[Knob-based pairing.]{\includegraphics[scale=0.3]{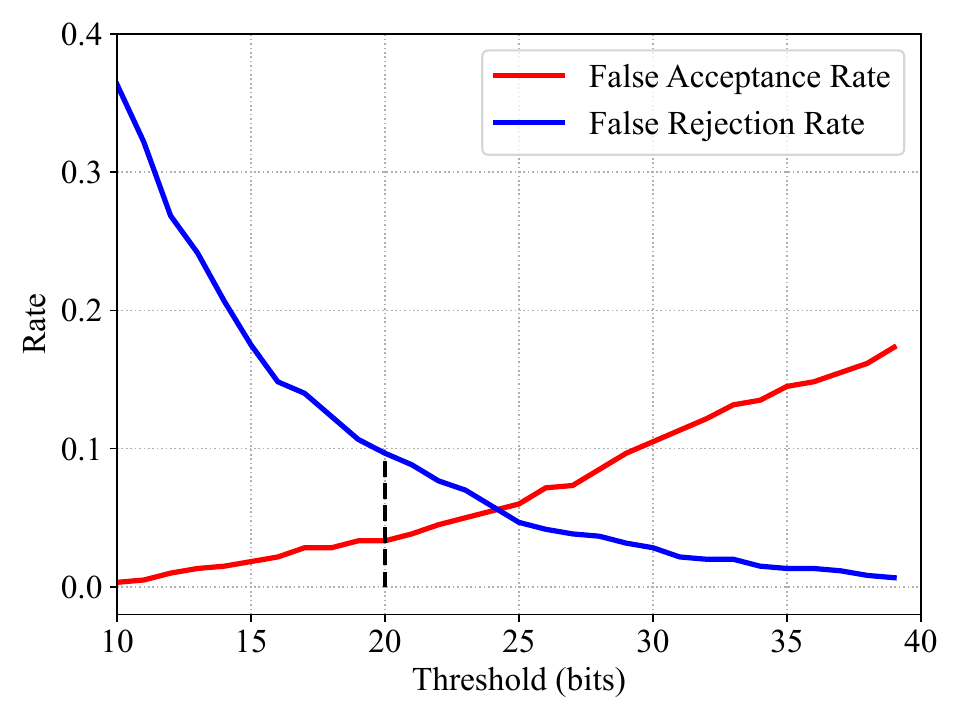}}~
\subfloat[Screen-based pairing.]{\includegraphics[scale=0.3]{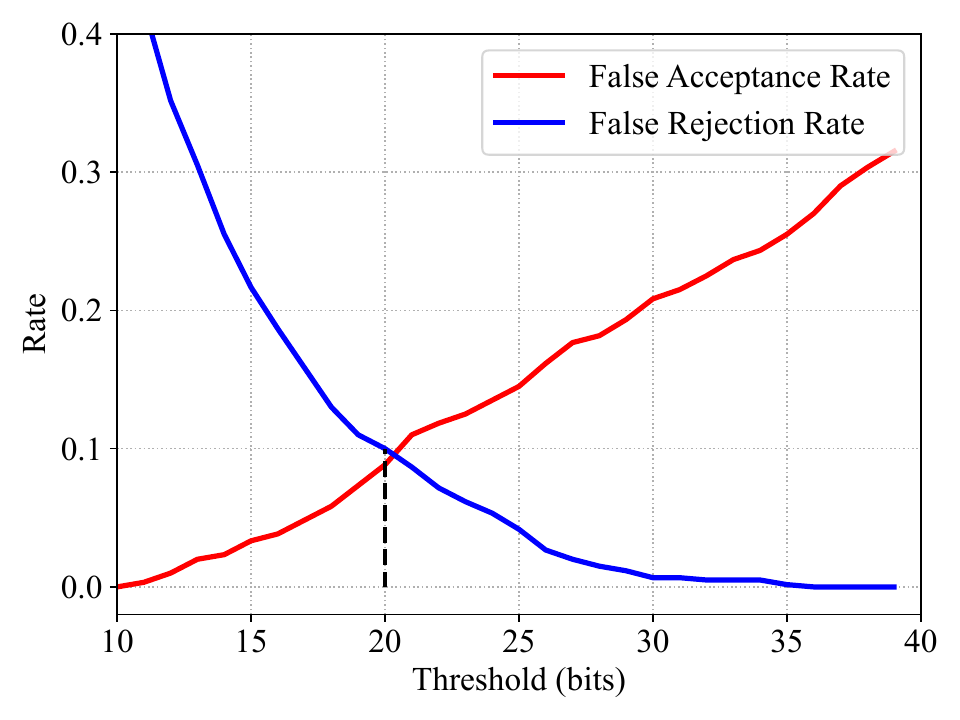}}~
\caption{\shortname: FARs and FRRs with different threshold values for pairing operations \emph{without} random pauses.}
\label{fig:pairing_performance_no_holding}
\aaf\aaf
\end{figure*}

\begin{figure*}[ht]
\centering
\subfloat[Button-based pairing.]{\includegraphics[scale=0.3]{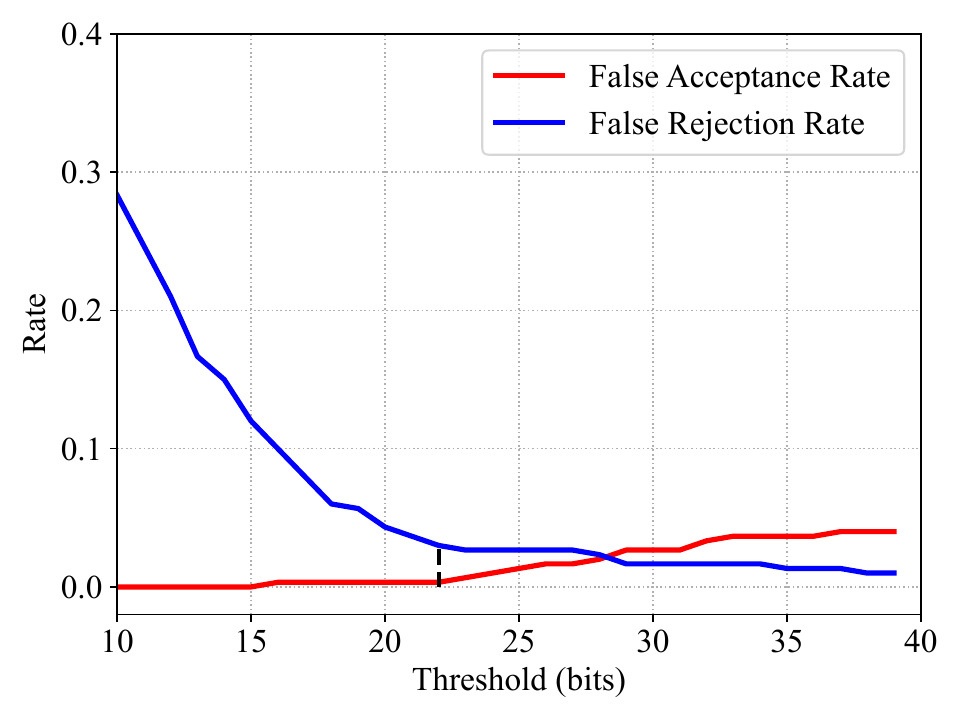}}~
\subfloat[Knob-based pairing.]{\includegraphics[scale=0.3]{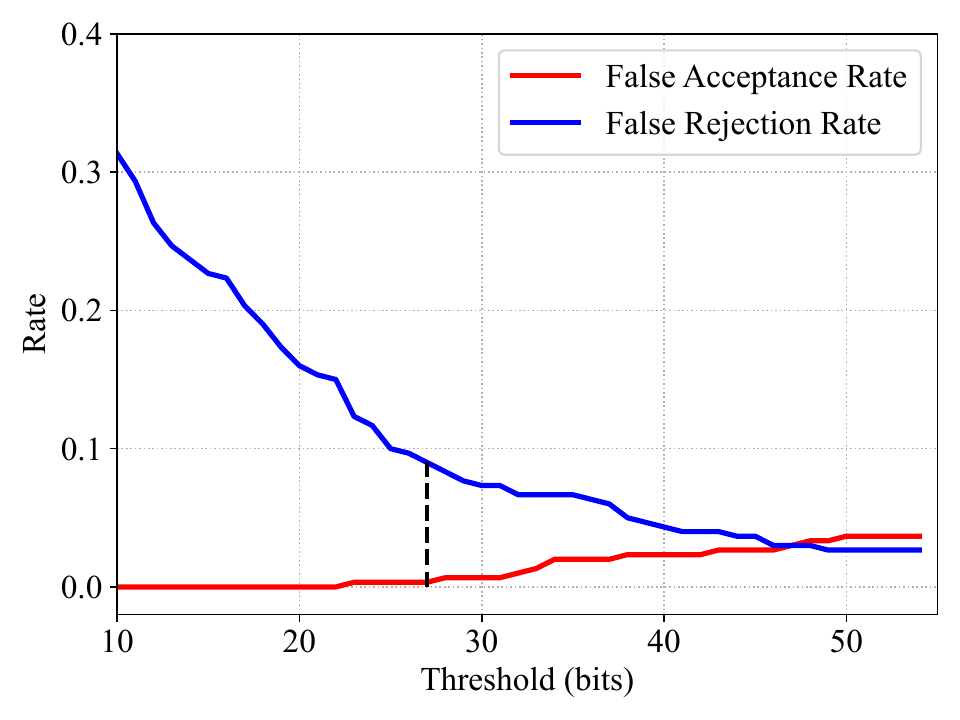}}~
\subfloat[Screen-based pairing.]{\includegraphics[scale=0.3]{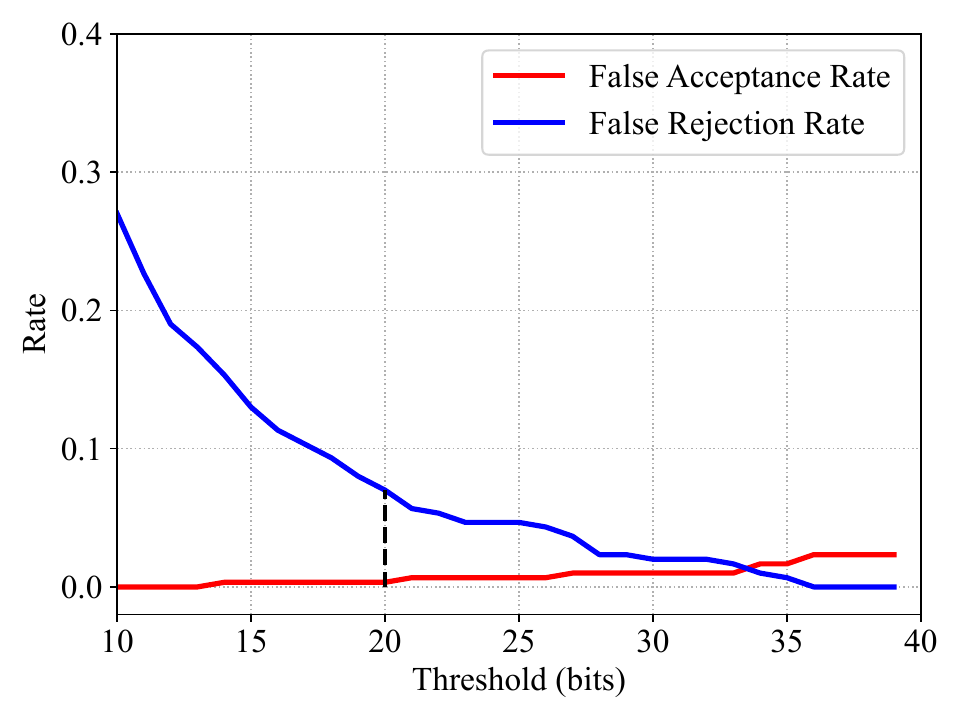}}~
\caption{\shortname: FARs and FRRs with different threshold values for pairing operations \emph{with} random pauses.}
\label{fig:pairing_performance}
\aaf\aaf
\end{figure*}

\subsection{Pairing Accuracy}
\label{subsec:performance}

We utilize Dataset I to evaluate the accuracy of our system and compare the performance between pairing operations with and without pauses. We employ the \emph{False Rejection Rate} (FRR) and \emph{False Acceptance Rate} (FAR) as metrics for measuring pairing accuracy.
1) FRR represents the rate at which our system fails to pair the legitimate user's IoT device with the helper. Achieving a low FRR is crucial for usability.
2) FAR denotes the rate at which our system pairs the legitimate user's IoT device (resp. helper) with the attacker's helper (resp. IoT device), respectively. Thus, maintaining a low FAR is paramount for security.

\vspace{2pt}
\noindent
\textbf{\faithful}
Given a pairing operation, \faithful accepts the pairing as legitimate if a shared key can be successfully derived from evidence pairs with a Hamming distance smaller than the threshold ($Thr$), as detailed in Section~\ref{subsec:challenges_and_solutions}. A false rejection occurs when $\mathrm{Ham}(E_{d_1}, E_{d_2}) > {Thr}$ for a legal pairing of $d_1$ and $d_2$, while a false acceptance happens if $\mathrm{Ham}(E_{d_1}, E_{d_3}) < {Thr}$ for an illegal pairing of $d_1$ and $d_3$. The evidence length denotes the number of time intervals it contains, set to 7 for knobs and 6 for both touchscreens and buttons in pairings with pauses (see \textbf{\emph{Evidence Length}} in Section~\ref{subsec:stability}). For pairings without pauses, the evidence length is uniformly set to 8 for all devices.

Figure~\ref{fig:pairing_performance_no_holding} illustrate the performance, without pauses, regarding FAR and FRR across varying Hamming distance thresholds. We establish the base value as 10ms (\textbf{\emph{Base Value}} is  studied in Section~\ref{subsec:stability}). Generally, higher thresholds yield lower FRR (better usability) but higher FAR (reduced security). By selecting a threshold resulting in an FRR of 0.10 (deemed reasonably good for usability), we achieve an FAR of 0.02, 0.03, and 0.09 for buttons, knobs, and touchscreens, respectively (indicated by \emph{vertical dashed lines}). An FRR of 0.10 means that, on average, 10 out of 100 pairing attempts fail, and thus a user is expected to perform $100/90=1.1$ pairing attempts for pairing one device.

Figure~\ref{fig:pairing_performance} presents performance when \emph{random pauses} are introduced during pairing. Notably, FAR can be significantly reduced, with its growth rate slowing as the threshold increases.   For applications prioritizing security, FAR can be set to 0.00, resulting in (FAR, FRR) of (0.00, 0.03) for buttons, (0.00, 0.09) for knobs, and (0.00, 0.07) for touchscreens (marked by \emph{vertical dashed lines}). This indicates that \emph{pauses enhance the accuracy for all the three types of UIs}. Thus, the rest of the evaluation is focused on the design with pauses.

In contrast, when vanilla fuzzy commitment (Section~\ref{subsec:challenges_and_solutions})  is used, we observe significantly degraded accuracies, with (FAR, FRR) of (0.00, 0.81) for buttons, (0.00, 0.48) for knobs, and (0.00, 0.73) for touchscreens.

\vspace{2pt}
\noindent
\textbf{\precise}
Given a pairing operation, \precise accepts the pairing if the machine learning classifier determines that the evidence pair is correlated.  We set the evidence length the same as that in \faithful for comparison. We employ a machine learning classifier, leveraging an 80\% portion of Dataset I for training, while reserving the remaining 20\% for testing. Our evaluation includes support vector machine (SVM), k-nearest neighbors (k-NN), and random forest (RF), with SVM demonstrating the highest performance.

\begin{table}
    \small
    \centering
    \caption{\faithful vs.\ \precise: AUCs and EERs}
    \af 
    \renewcommand{\arraystretch}{1.05}
    
    \begin{tabular}{c|c|c|c|c}
	\hline
    System &  Pauses? &  Device  &  AUC & EER \\\hline
    
    \multirow{6}{*}{\faithful}  &  \multirow{3}{*}{No}  &  Button  &  0.9939 & 0.0320 \\
    & & Knob & 0.9872 & 0.0560 \\
    & & Screen & 0.9761 & 0.0956 \\ \cline{2-5}
    & \multirow{3}{*}{Yes}  &  Button  &  0.9940 & 0.0217 \\
    & & Knob & 0.9972 & 0.0300 \\
    & & Screen & 0.9990 & 0.0133 \\ \hline
    \multirow{6}{*}{\precise}  &  \multirow{3}{*}{No}  &  Button  &  0.9982 & 0.0250 \\
    & & Knob & 0.9956 & 0.0233 \\
    & & Screen & 0.9937 & 0.0400 \\ \cline{2-5}
    & \multirow{3}{*}{Yes}  &  Button  &  0.9997 & 0.0133 \\
    & & Knob & 0.9993 & 0.0167 \\
    & & Screen & 0.9995 & 0.0117 \\ \hline
    \end{tabular}
\label{tab:auc_eer}
\aaf\aaf
\end{table}

\begin{figure}
\centering
\includegraphics[scale=0.3]{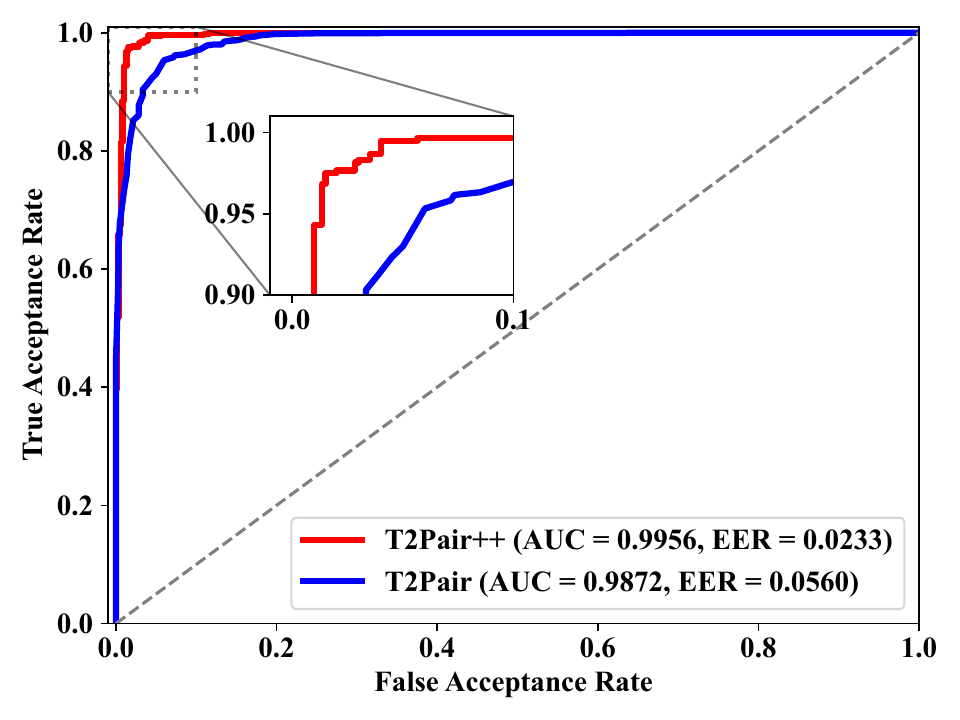}
\aaf
\caption{\shortname vs.\ \precise: ROC curves \emph{without} random pauses for Knob-based device pairing.}
\label{fig:pairing_auc_without_pause}
\aaf\aaf\aaf
\end{figure}

\begin{figure}
\centering
\includegraphics[scale=0.3]{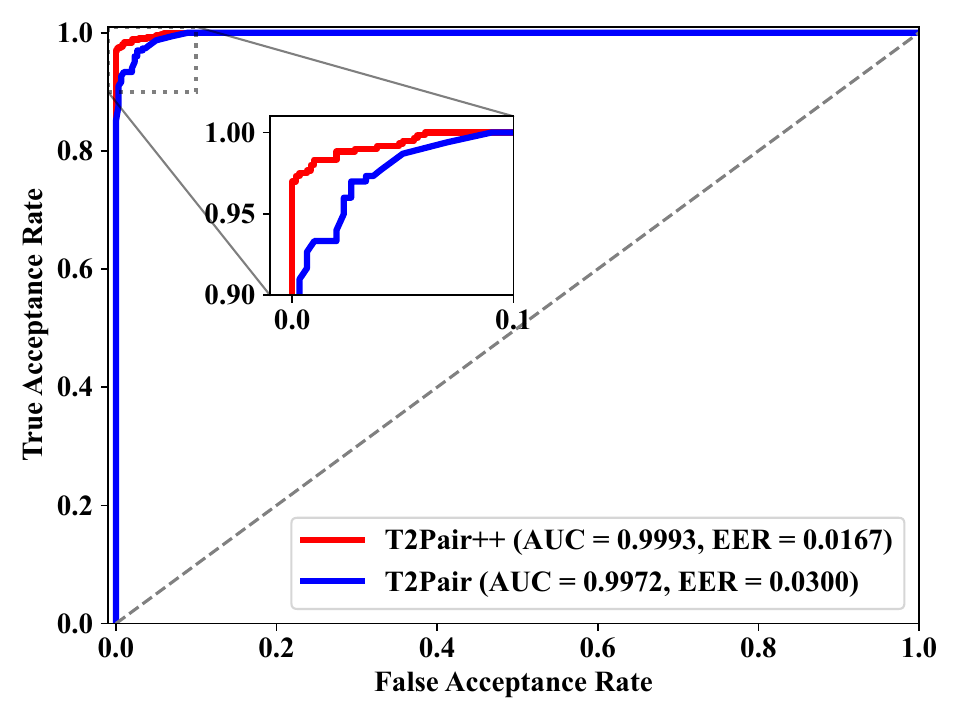}
\aaf
\caption{\shortname vs.\ \precise: ROC curves \emph{with} random pauses for Knob-based device pairing.}
\label{fig:pairing_auc_with_pause}
\aaf\af
\end{figure}

We compare the Area Under the Curve (AUC) values and Equal Error Rate (EER) associated with the two protocols. AUC is a performance metric of a system's ability to discriminate between classes across different threshold values, while EER denotes the point at which the FAR equals FRR. Table~\ref{tab:auc_eer} shows the results. With pauses enabled, \precise attains AUCs of 0.9997, 0.9993, and 0.9995 for buttons, knobs, and touchscreens, while \faithful achieves 0.9940, 0.9972, 0.9990, respectively. In terms of EERs, \precise achieves lowers rates of 0.0133, 0.0167, and 0.0117 for buttons, knobs, and touchscreens, whereas \faithful achieves rates of 0.0217, 0.0300, and 0.0133, respectively. As an example, Figure~\ref{fig:pairing_auc_without_pause} and Figure~\ref{fig:pairing_auc_with_pause} compare the AUCs and EERs achieved by \faithful and \precise in the scenario of knob-based device pairing with and without pauses, respectively. \precise consistently achieves higher AUCs than \faithful and lower EERs, regardless of whether pauses are introduced or not. Furthermore, when pauses are enabled, both protocols achieve higher AUCs and lower EERs compared to when pauses are not enabled, demonstrating the effectiveness of adding intentional pauses during pairing operations.

In summary, our system demonstrates improved accuracy when pauses are enabled compared to when they are not. Additionally, \precise consistently outperforms \faithful in terms of pairing accuracy across buttons, knobs, and touchscreens, whether pauses are enabled or not.

\subsection{Resilience to Mimicry Attacks}
\label{subsec:security}
This section evaluates the resilience of our system (based on the thresholds that result in a FRR of 0.10 as discussed in Section~\ref{subsec:performance}) to mimicry attacks for two types of pairing operations: one without pauses (\emph{Type-I}) and the other with pauses (\emph{Type-II}). We use FAR to measure the success rate of attacks. We evaluate the resilience using Dataset II (see Section~\ref{subsec:dataset2}).

\begin{table}
    \centering
    \small
    \renewcommand{\arraystretch}{1.0}
    \caption{FARs under mimicry attacks.} 
    \label{tab:attack_accuracy1}
        \aaf\af
    \resizebox{0.48\textwidth}{!}{
    \begin{tabular}{c|c|c|c|c}
    \hline
    \multirow{2}{*}{Attacks}  &  \multirow{2}{*}{Pauses?}  &  \multirow{2}{*}{Device}  & \multicolumn{2}{c}{Average FAR}\\  
    \cline{4-5}
    & & & \faithful & \precise \\ \hline
    
    \multirow{6}{*}{\begin{tabular}{@{}c@{}} \textbf{MA-obstructed} \\ No clear view \\ Untrained attackers \end{tabular}}  & \multirow{3}{*}{No}
    & Button   & 0.040 & 0.027\\
    & &  Knob  & 0.047 & 0.033\\
    & & Screen & 0.093 & 0.047\\ \cline{2-5}
    
    &  \multirow{3}{*}{Yes}
    & Button & 0.007 & 0.000\\
    & & Knob & 0.000 & 0.000\\
    & & Screen & 0.007 & 0.000\\ \hline
    \multirow{6}{*}{\begin{tabular}{@{}c@{}} \textbf{MA-clear}  \\ Clear view  \\ Untrained attackers \end{tabular}}  & \multirow{3}{*}{No}    
    & Button & 0.093  & 0.053\\ 
    & & Knob & 0.100  & 0.053\\
    & & Screen & 0.180  & 0.087\\ \cline{2-5}
    
    &  \multirow{3}{*}{Yes}
    & Button & 0.020 & 0.000\\ 
    & & Knob   & 0.040 & 0.020\\
    & & Screen & 0.020 & 0.013\\\hline
    \multirow{6}{*}{\begin{tabular}{@{}c@{}} \textbf{MA-trained}  \\ Clear view  \\ Trained attackers  \end{tabular}}  &  \multirow{3}{*}{No}  &  Button  & 0.274 & 0.213 \\
    &   &  Knob  & 0.240 & 0.187 \\
    &   &  Screen  & 0.180 & 0.120 \\ \cline{2-5}
    
    &  \multirow{3}{*}{Yes}  
    & Button  & 0.040 & 0.027 \\ 
    &  &  Knob  & 0.040 & 0.020 \\ 
    &  &  Screen  & 0.027 & 0.013 \\ \hline
    
    \end{tabular}}
    \af
\end{table}

\vspace{2pt}
\noindent\textbf{Resilience to \textbf{MA-obstructed}.} 
The attacker ($\mathcal{A}$) stands behind the victim ($\mathcal{V}$) with a distance of 2--3 meters and does not have a clear view of $\mathcal{V}$'s hand movements. 
As shown in Table~\ref{tab:attack_accuracy1}, 
for the pairing operations of \emph{Type-I}, \faithful can successfully identify 96.0\%, 95.3\% and 90.7\% of attacks on buttons, knobs, and touchscreens, respectively. \precise can successfully identify 97.3\%, 96.7\% and 93.3\% of attacks on buttons, knobs, and touchscreens, respectively, which is higher than that of \faithful. 
The performance can be greatly improved if the random pauses are considered. Specifically, for the \emph{Type-II} operations, \faithful can successfully defend against all the attacks on knobs, and 99.3\% of attacks on touchscreens and buttons, while \precise can identify all the attacks.

\vspace{2pt}
\noindent\textbf{Resilience to \textbf{MA-clear}.} 
$\mathcal{A}$ stands next to $\mathcal{V}$ and  
has a clear view of $\mathcal{V}$'s hand movements. 
As shown in Table~\ref{tab:attack_accuracy1}, 
for the \emph{Type-I} operations, the attackers' success rate increases, no matter under \faithful or \precise. 
However, for the \emph{Type-II} operations, the attackers' success rate is still very low for both protocols. 
The results demonstrate that the random pauses during each pairing can increase the difficulty for attackers to mimic the victims' hand movements. 
Thus, the pairing operations with random pauses are more secure. Moreover, Table~\ref{tab:attack_accuracy1} shows that \precise achieves lower FARs than \faithful.

\vspace{2pt}
\noindent\textbf{Resilience to \textbf{MA-trained}.}
How to train the attacker is described in Section~\ref{subsec:dataset2}).
Compared to the \emph{Type-II} operations, as shown in Table~\ref{tab:attack_accuracy1}, FARs for the \emph{Type-I} operations increase sharply (up to 27.4\% under \faithful and 21.3\% under \precise),
which reveals a noticeable \textbf{\emph{weakness}} of pairing without pauses.
The pauses make the intervals more unpredictable and difficult to mimic, as the FARs remain low when pauses are enabled, regardless of \faithful or \precise.
To eliminate the weakness, both protocols perform self-checking at the phase of Extracting Evidence, which aborts pairing if there are no pauses. It can also be observed that, compared to \faithful, \precise has lower FARs for MA-trained attacks.

In sum, the resilience of our system to mimicry attacks can be enhanced by adding intentional pauses during pairing operation. Moreover, \precise demonstrates superior resilience against mimicry attackers compared to \faithful. For instance, with pauses enabled, \precise achieves lower FARs of 0.027, 0.020, and 0.013 for buttons, knobs, and touchscreens under \textbf{MA-trained}, as opposed to 0.040, 0.040, and 0.027 achieved by \faithful.

\begin{table}
    \small
    \centering
    \caption{NIST statistical test results. A $p$-value greater than 0.01 indicates a randomness test is passed.}
    \af 
    \renewcommand{\arraystretch}{1.05}
    
    \begin{tabular}{c|c|c|c}
	\hline
    \multirow{2}*{Test}  &  \multicolumn{3}{c}{$p$-value}  \\\cline{2-4}
    &   Button  &  Knob  &  Screen \\\hline
    
    Frequency  &  0.327  &  0.581  &  0.300\\\hline
    
    Block Frequency  &  0.854  &  0.118  &  0.807\\\hline
    
    Runs  &  0.190  &  0.697  &  0.046 \\\hline
    
    Longest Run  &  0.249  &  0.624  &  0.164 \\\hline
    
    Approximate Entropy  &  0.051  &  0.369  &  0.095\\\hline
    
    FFT  &  0.567  &  0.567  &  0.829\\\hline
    
    Cumulative Sums (Fwd)  &  0.537  &  0.318  &  0.505\\\hline
    
    Cumulative Sums (Rev) &  0.476  &  0.681  &  0.343\\\hline
    
    \multirow{2}*{Serial}  &  0.387  &  0.251  &  0.360 \\\cline{2-4}
    &  0.601  &  0.074  &  0.796 \\\hline
    \end{tabular}
\label{tab:nist_test}
\af
\end{table}

\subsection{Randomness and Entropy}
\label{subsec:randomness_entropy}

\noindent\textbf{Randomness.}
The randomness level of the time interval between two consecutive events directly impacts the entropy of evidence. 
Evaluating the randomness of these intervals poses a challenge due to the requirement for a large number of samples. Prior work~\cite{shakeWell, Perceptio} has also acknowledged this challenge and directly assumes that human-generated events are random.
Similar to H2H~\cite{H2H_CCS14}, we investigate whether the six least significant bits of the time intervals are randomly distributed. To verify this, we apply the NIST statistical test suite~\cite{nist} to the distribution of our time interval bits. This test suite is widely used for assessing randomness in various contexts~\cite{H2H_CCS14, TDS_ccs16, touch_and_guard_tmc}. Our dataset, which is subsampled from Dataset I and II based on users, comprises 19.2 Kbits, consisting of 3200 intervals for each type of pairing operation.

The outputs of the NIST tests are $p$-values, where each $p$-value represents the probability that the input bit sequence is generated by a random bit generator~\cite{nist}. If a $p$-value is less than a chosen critical value (usually 0.01), the null hypothesis for randomness is rejected. Table~\ref{tab:nist_test} demonstrates that all the $p$-values for the three types of devices are greater than 0.01. These results confirm the randomness of the collected time intervals.

\vspace{2pt}
\noindent\textbf{Entropy analysis.}
We denote the set of intervals generated without pauses as $I_1$, and the set of intervals with pauses as $I_2$. The possible range of $I_1$ is related to the specifications of a given device (\eg size, rotation/swiping range) and the behavior habits of device users. On the other hand, the range of $I_2$ is mainly determined by the behavior of device users.

As many human characteristics exhibit normal distributions~\cite{borghi1965distribution}, we assume that $I_1$ and $I_2$ among all users follow a normal distribution each. The entropy (in bits) of a time interval, with mean denoted as $\mu$ and standard deviation as $\sigma$, can be computed as follows~\cite{information_book_93}.

\begin{equation}
\thickmuskip=0.5\thickmuskip
\begin{split}
    E_i 
    = \frac{1}{2} \log_{2} (2\pi e \sigma^2)
\end{split}
\label{eq:entropy}
\aaf
\end{equation}
Assuming each piece of evidence contains $n_1$ intervals from $I_1$ and $n_2$ intervals from $I_2$, the entropy of the evidence can be computed as:
\begin{equation}
    l_E = n_1 * E_1 + n_2 * E_2 + \log_2 \binom{n_1 + n_2}{n_2}
\end{equation}
\af
The term $\binom{n_1 + n_2}{n_2}$ is introduced to account for the random occurrence positions of the $n_2$ pauses in the evidence.

The total time of generating a piece of evidence is denoted as $T$. Then, the bit rate is given by ${l_E}/{T}$.

\begin{figure}
\aaf
\subfloat[Intervals of $I_1$.]{\includegraphics[scale=0.268]{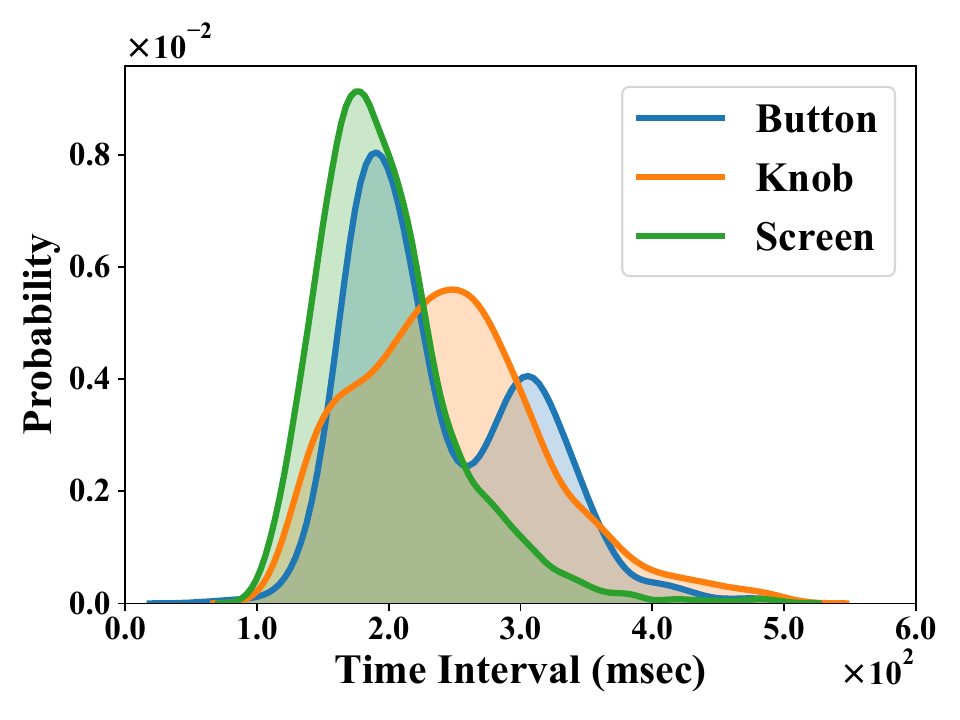}}
~
\subfloat[Intervals of $I_2$.]{\includegraphics[scale=0.268]{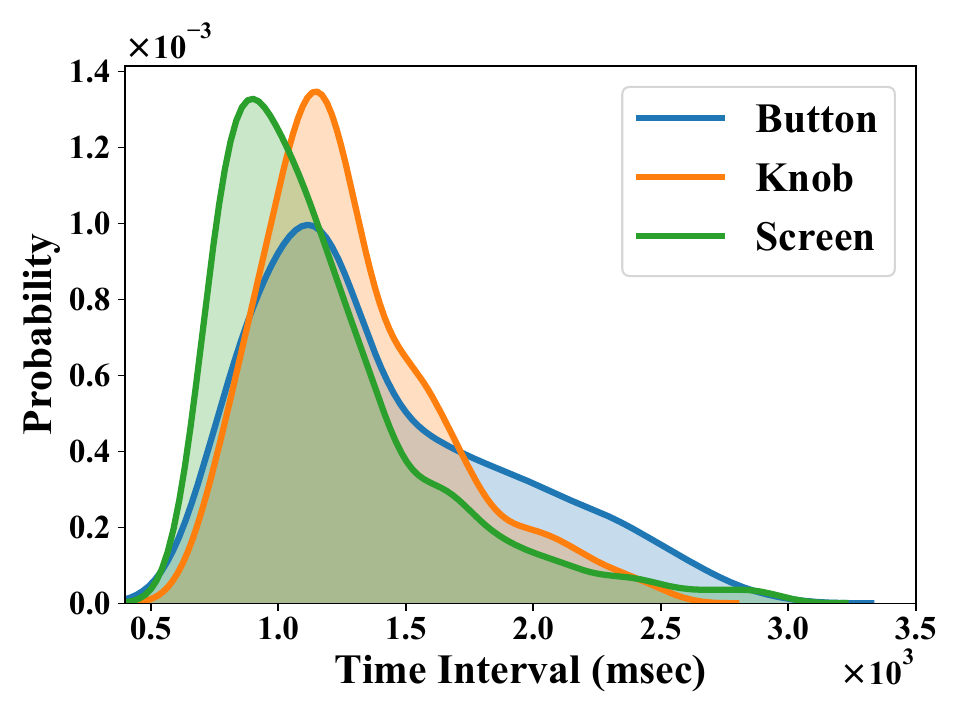}}
\af 
\caption{Time interval distributions.}
\aaf
\label{fig:distribution}
\end{figure}

\vspace{2pt}
\noindent\textbf{Entropy evaluation using a real-world dataset.}
Figure~\ref{fig:distribution} illustrates the distributions of the time intervals of $I_1$ and $I_2$ among all users. We test the normality of these distributions using one-sample Kolmogorov-Smirnov testing~\cite{KL_test}. For each device, more than 86\% of the time intervals conform to the normality assumption. Therefore, most of the data for each device can be approximated by a normal distribution. This finding is consistent with prior studies~\cite{keystroke_observ, button_press} on keystrokes and/or screen touches.

We use the pairing operations on buttons as a case study to compute the entropy. As summarized in Table~\ref{tab:entropy}, the intervals of $I_1$ mostly fall in the range of [100ms, 500ms] with a standard deviation $\sigma_1$ of 67ms, while those of $I_2$ are distributed in the range of [800ms, 3000ms] with a standard deviation $\sigma_2$ of 501ms. In \faithful, with the base value $=10$ms (see Section~\ref{subsec:stability}), $\sigma_1$ and $\sigma_2$ become 6.7 and 50.1, respectively. According to our entropy definition in Equation~\ref{eq:entropy}, the entropy for one interval in $I_1$ is approximately 4.8 bits, and that in $I_2$ is around 7.7 bits. As each piece of evidence consists of 4 (or 5) intervals of $I_1$ and 2 (or 1) intervals of $I_2$, the total entropy is around 38.5 (or 34.3) bits. The mean values for the intervals of $I_1$ and $I_2$ are 238ms and 1402ms, respectively. Therefore, the total time for generating a piece of evidence is 3756ms (or 2592ms). The bit rate is around 10.3 bit/s (or 13.2 bit/s).

By contrast, in \precise, as the intervals are not divided by any base value, the $\sigma_1$ remains 67 and $\sigma_2$ remains 501. The entropy for one interval in $I_1$ is around 8.1 bits, and that in $I_2$ around 11.0 bits. Thus, the total entropy is around 54.4 (or 51.5) bits, and the bit rate is around 14.4 bit/s (or 19.7 bit/s).

In short, \precise achieves higher entropy and entropy rate compared to \faithful. For instance, when 2 pauses are enabled, \precise achieves an entropy of 54.5, surpassing \faithful's 38.5, and an entropy rate of 14.4 bit/s, outperforming \faithful's 10.3 bit/s.

\begin{table}[t]
    \small
    \centering
    \caption{Average entropy and estimated bit rate.}
    \aaf 
    \renewcommand{\arraystretch}{1.05}
    
    \begin{tabular}{c|c|c|c}
	\hline
      &  Button  &  Knob  &  Screen \\\hline
    
    $\sigma$ of $I_1$ (ms)  &  67  &  72  &  53 \\\hline
    
    Entropy of $I_1$ in T2Pair (bits)  &  4.79  &  4.90  &  4.45 \\\hline
    
    Entropy of $I_1$ in Our work (bits)  &  8.11  &  8.21  &  7.78 \\\hline
    
    $\sigma$ of $I_2$ (ms)  &  501  &  362  &  424 \\\hline
    
    Entropy of $I_2$ in T2Pair (bits)  &  7.69  &  7.23  &  7.45\\\hline
    
    Entropy of $I_1$ in Our work (bits)  &  11.01  &  10.55  &  10.78 \\\hline
    \end{tabular}
\label{tab:entropy}
\aaf\aaf
\end{table}

\begin{figure*}
\aaf\aaf\af
\centering
\subfloat[FAR \emph{vs.} evidence length]{\includegraphics[scale=0.3]{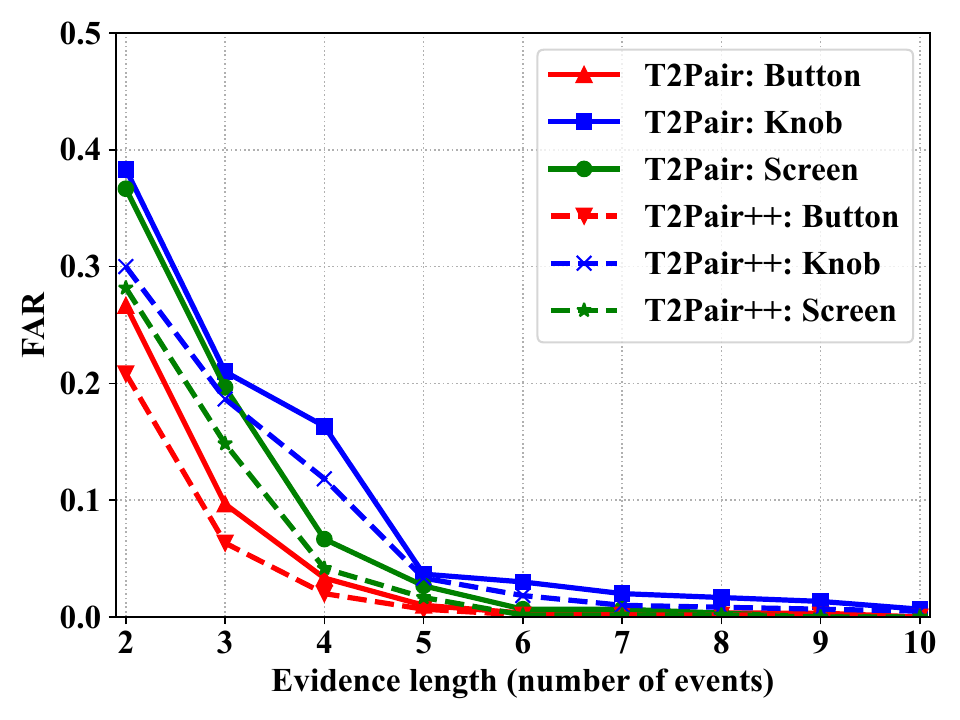}}
\subfloat[EER \emph{vs.} base value]{\includegraphics[scale=0.3]{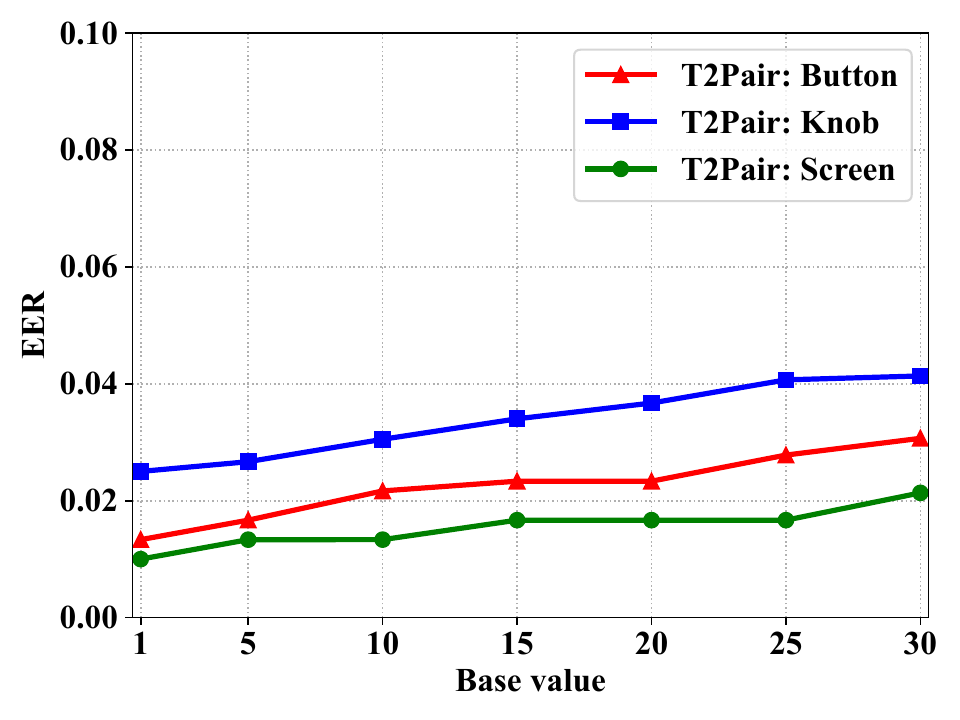}}
\subfloat[EER \emph{vs.} sampling rate]{\includegraphics[scale=0.3]{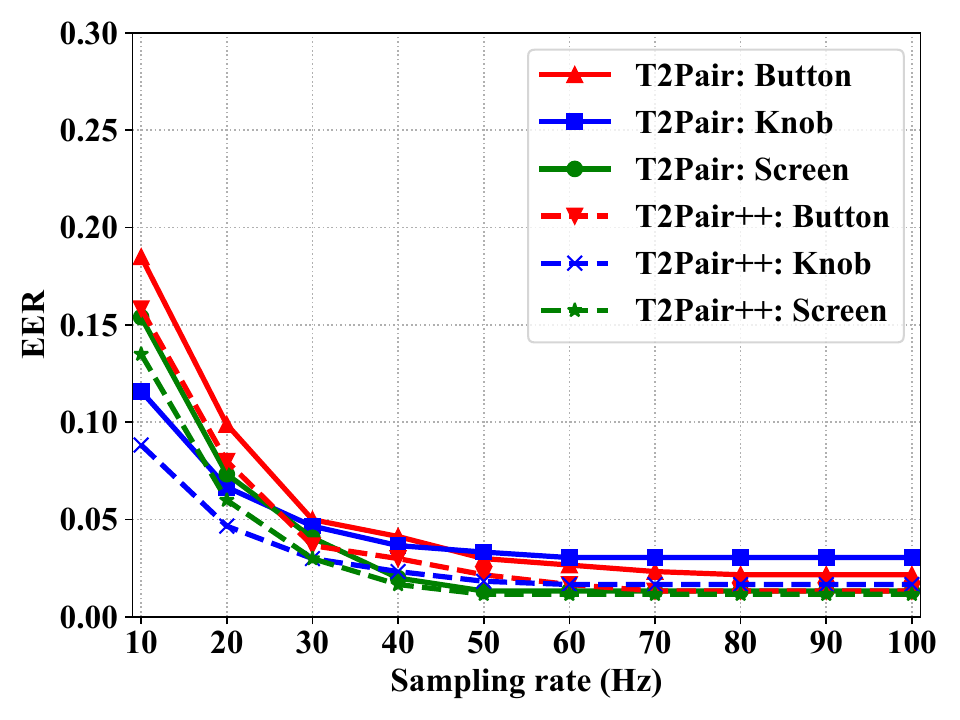}} \\
\aaf
\subfloat[EER \emph{vs.} device position]{\includegraphics[scale=0.3]{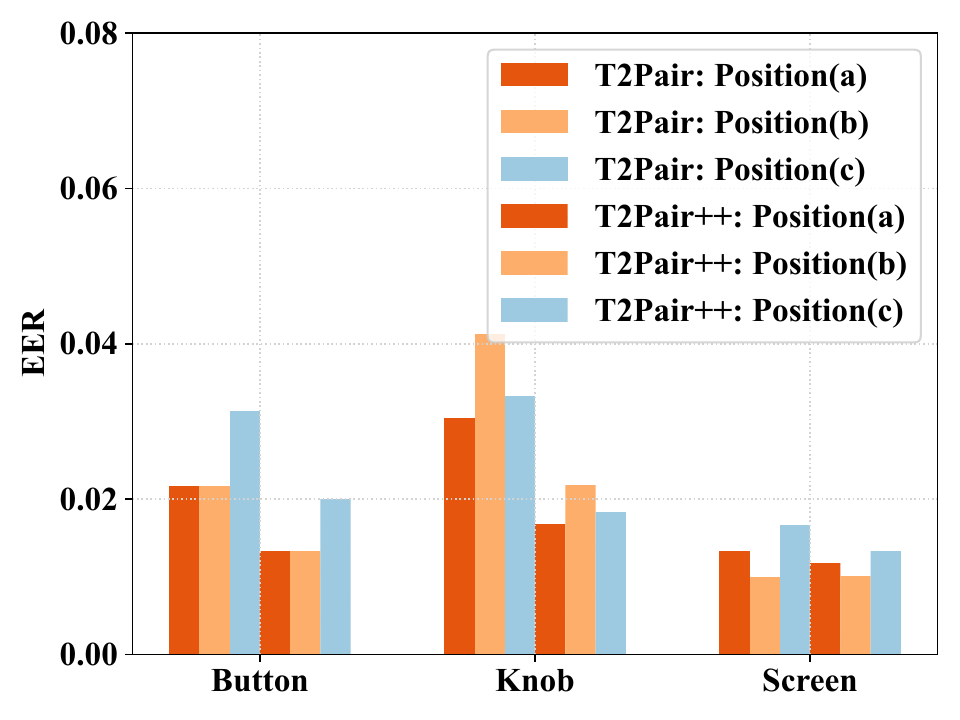}}
\subfloat[EER \emph{vs.} helper]{\includegraphics[scale=0.3]{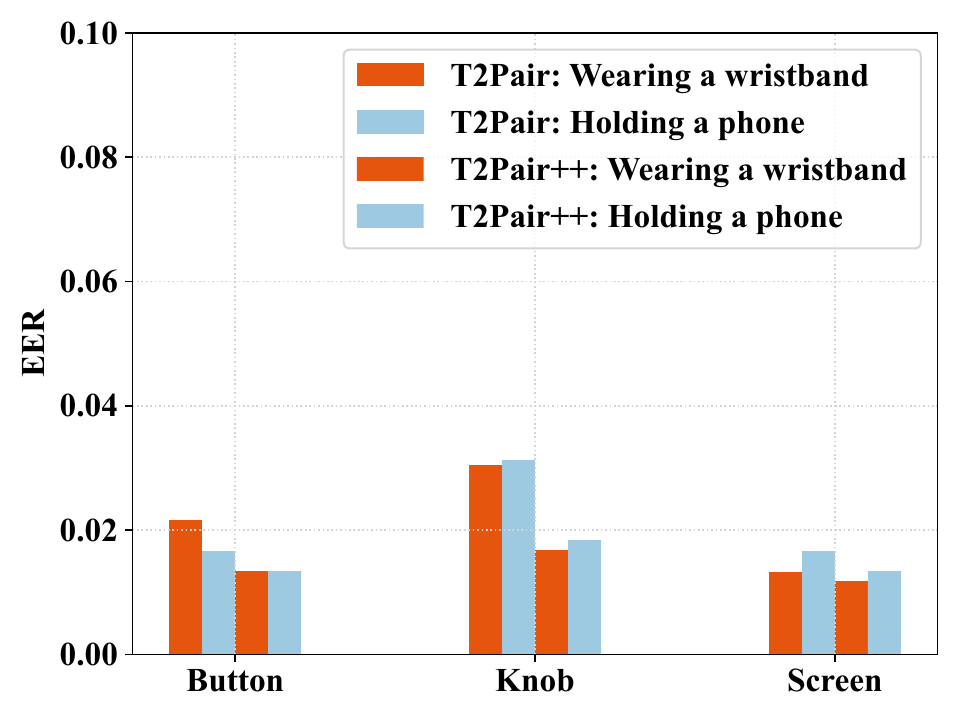}}
\subfloat[EER \emph{vs.} IoT device]{\includegraphics[scale=0.3]{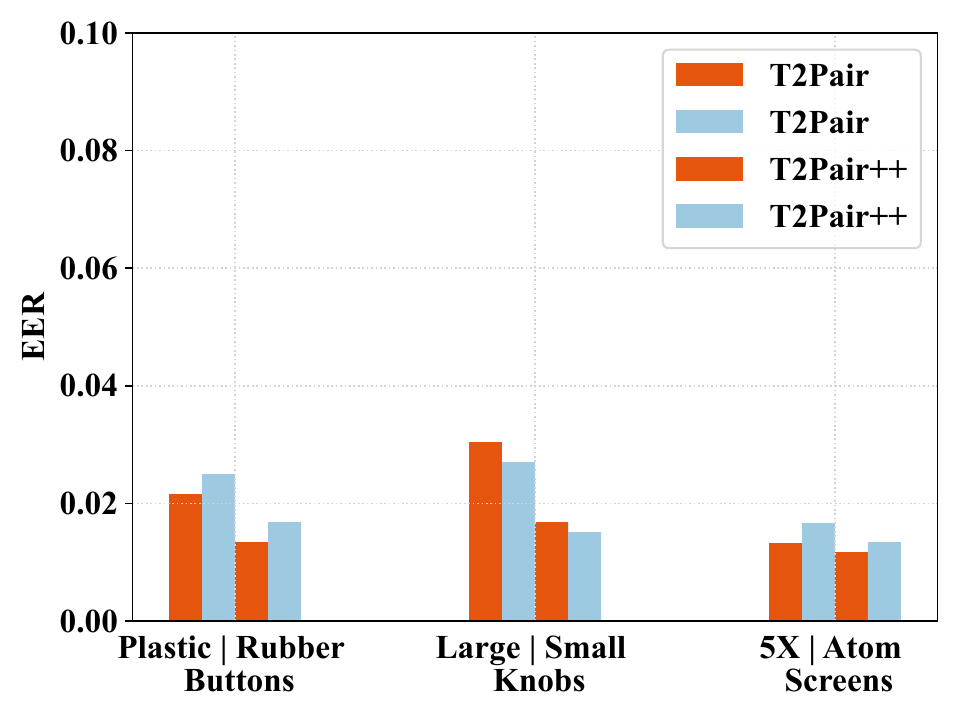}}
\af
\caption{Impacts of different parameters and experimental settings.}
\label{fig:parameter}
\aaf\aaf
\end{figure*}

\subsection{Study of Parameters and Stability}
\label{subsec:stability}
For the following experiments, we focus on pairing with pauses.

\vspace{2pt}
\noindent \textbf{Evidence length.}
The evidence length is denoted by the number of time intervals, which correlates with the number of salient points. Longer evidence typically translates to improved security but comes at the cost of increased pairing time, thereby affecting usability. Consequently, determining the appropriate evidence length involves striking a balance between security and usability. It is also worth noting that the evidence length can be adjusted based on the desired balance between security and usability.

To investigate its influence, we fix the FRR at 0.05 and analyze how the FAR changes with varying evidence lengths. Figure~\ref{fig:parameter}(a) illustrates the FARs with different evidence lengths for the three device types. As anticipated, longer evidence lengths lead to lower FARs, signifying enhanced security.
In \faithful, for knob-based devices, an evidence length of 7 is deemed suitable, as further increasing the length marginally improves the FAR. For button-based and screen-based devices, the FAR drops below 0.01 with evidence lengths exceeding 6. Thus, 6 is considered an appropriate length for these devices.
Moreover, compared to \faithful, \precise achieves lower FARs with same evidence lengths. For the sake of comparison, we maintain consistent evidence lengths for both protocols.

\vspace{2pt}
\noindent \textbf{Base value.}
The base value plays a crucial role in encoding time intervals. While a larger base may lead to less accurate representations due to coarse approximations, it can result in shorter evidence encodings, enhancing efficiency. Hence, selecting the appropriate base value involves balancing accuracy and efficiency. We employ the EER to evaluate the impact of the base value, giving equal weight to FAR and FRR.

Figure~\ref{fig:parameter}(b) displays the EERs for the three pairing types across various base values ranging from 1 to 30ms. Notably, we observe a gradual increase in EERs with higher base values. Although a base smaller than 10ms may slightly enhance EERs, it also results in longer evidence encodings. Considering both accuracy and efficiency, we choose the base value as 10ms.

\vspace{2pt}
\noindent \textbf{Sampling rate.}
The sensor data from the wristband 
is used to extract salient points and generate the evidence. A low sampling rate of the sensor data may lead to inaccuracies in detecting salient points, while a high sampling rate can capture more subtle motions but imposes a greater burden on data collection. Therefore, determining the optimal sampling rate involves balancing accuracy and efficiency.

Figure~\ref{fig:parameter}(c) illustrates the performance 
across varying sampling rates from 10Hz to 100Hz in increments of 10Hz. We observe that, irrespective of \faithful or \precise, button clicking necessitates a sampling rate higher than 80Hz to achieve optimal performance, whereas knob rotation and screen swiping require a sampling rate higher than 50Hz. Consequently, we select the sampling rate of 80Hz.


\vspace{1pt}
\noindent \textbf{IoT device position.}
IoT devices may be installed or placed in various positions based on demand, such as the need for a power source, or user preference. We examine three common positions of IoT devices: (a) plugged into a wall outlet, (b) placed on a table, and (c) held in a hand.

Figure~\ref{fig:parameter}(d) illustrates the EERs 
across different device positions. Comparable observations are evident across both protocols. For buttons and touchscreens, our system performs slightly better when devices are positioned on a wall outlet or a table, while for knobs, performance is slightly better when devices are held in hand or plugged into a wall outlet. Overall, the results suggest that different device positions have minimal impact on pairing performance.

\vspace{1pt}
\noindent \textbf{Different kinds of helpers.}
We investigate the feasibility of using a smartphone 
for performing pairings. In Figure~\ref{fig:parameter}(e), we present the EERs for the three types of pairings using either the wristband and the smartphone as the helper. Across all protocols, we observe no significant difference in pairing performance between the two helpers. Therefore, we conclude that holding a smartphone for pairings is indeed feasible. However, we notice that usability is not satisfactory when the user holds a smartphone to twist a small knob.

\vspace{1pt}
\noindent \textbf{Different sizes and materials of IoT devices.}
In Figure~\ref{fig:parameter}(f), we examine whether our system can effectively pair with IoT devices of varying sizes and materials. We consider two knob-based devices (a large knob and a small knob), two button-based devices (a rubber keypad and a plastic keypad), and two touchscreens (a smartphone Nexus 5X and a Unihertz Atom with different screen sizes).

The EERs for the six devices are shown in Figure~\ref{fig:parameter}(f). Remarkably, regardless of the protocol used, there is no significant difference observed in the performance between any two devices with the same type of user interface. Consequently, it can be inferred that the device size and material exert minimal impact on the pairing performance of our system.

\begin{figure}
\centering
\includegraphics[scale=0.3]{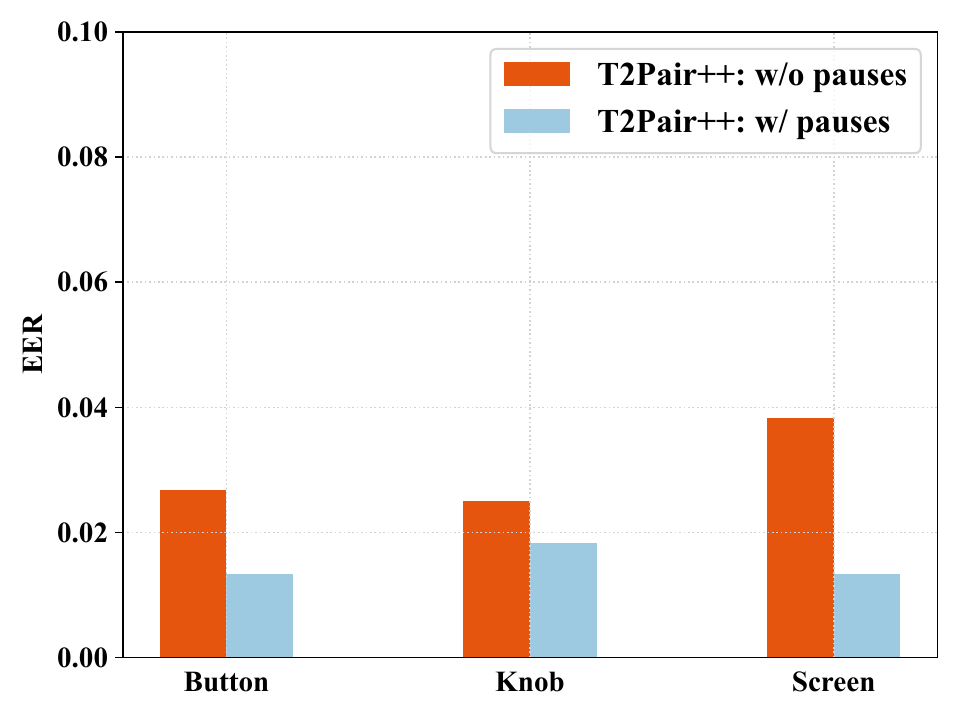}
\aaf
\caption{EERs evaluated using the LOSO validation strategy. The consistently low EERs demonstrate the system's ability to generalize to unseen users.}
\aaf\af
\label{fig:loso}
\end{figure}

\vspace{1pt}
\noindent \textbf{\cxr{Unseen Users.}} \cxr{To evaluate the system's performance on unseen users, we adopt a strict leave-one-subject-out (LOSO) cross-validation strategy. Specifically, we iteratively select one subject as the test case (i.e., unseen user) and train the system using data from the remaining 19 subjects. We then report the \emph{average} performance metrics across all iterations. This evaluation enables us to assess whether our system 
can be generalized to users not encountered during training. The results, based on the datasets described in Section~\ref{sec:data_collection}, are presented in Figure~\ref{fig:loso}. \precise consistently achieves low EERs, demonstrating strong generalization to unseen users.}

\subsection{Efficiency} \label{subsec:efficiency}
We evaluate the efficiency of the pairing operations, focusing solely on those with random pauses. Specifically, we measure \emph{the time used for performing the pairing operations} with an evidence length of 7 for knobs, 6 for touchscreens, and 6 for buttons (refer to \textbf{\emph{Evidence Length}} in Section~\ref{subsec:stability}).

For knobs, touchscreens, and buttons, the mean pairing time and Standard Deviation (SD) are 2.8s (SD=0.85), 2.3s (SD=0.66), and 3.2s (SD=0.93), respectively. These pairing operations require very little time to complete and are highly efficient.

For \faithful, we measure the time required to execute fuzzy commitment and PAKE to establish a shared key between two parties. The average execution time on the smartwatch and the Arduino controller is 0.9s (SD=0.37) and 0.7s (SD=0.25), respectively.

For \precise, the time used in the mutual authentication phase is measured. The average time taken in the mutual authentication phase on the smartwatch and the Arduino controller is 1.0s (SD=0.26) and 0.8s (SD=0.13), respectively.

\vspace{-1pt}
\subsection{Comparison with Other Approaches} \label{sec:comparison-other}

\begin{table}
\centering
\small
\caption{Comparison with other works.} 
\aaf\af
\renewcommand{\arraystretch}{1}
\begin{tabular}{l|c|c}
\hline
\textbf{Method}   &  \textbf{(FAR, FRR)}  &  \textbf{Time}(s)  \\\hline

\begin{tabular}{@{}l@{}}
ShaVe/ShaCK~\cite{shakeWell} \end{tabular}  &  (0.0, 0.10--0.12)  &  3 \\\hline

SFIRE~\cite{ghose18_infocom}  &  (0.0, -)  &  6  \\\hline

Tap-to-Pair~\cite{Tap-to-pair2018}  &  (-, 0.117)  &  15--20 \\\hline

Checksum~\cite{checksum_15ubicomp}  &  (-, 0.10)  &  5.7 \\\hhline{===}

\textbf{\faithful} & (0.0, 0.03--0.09)  &  3.2--4.1  \\  \hline 
\textbf{\precise} & (0.0, 0.02--0.03)  &  3.3--4.2  \\  \hline 
\end{tabular}
\label{tab:result_comparison}
\aaf\aaf 
\end{table}

Table~\ref{tab:result_comparison} illustrates the comparison of our system with several prior works. Our methods achieve superior accuracies compared to works such as ShaVe/ShaCK~\cite{shakeWell}, Tap-to-Pair~\cite{Tap-to-pair2018}, and Checksum~\cite{checksum_15ubicomp}. Additionally, \faithful and \precise demonstrates higher efficiency than Tap-to-Pair~\cite{Tap-to-pair2018}, SFIRE~\cite{ghose18_infocom}, and Checksum~\cite{checksum_15ubicomp} in terms of pairing time. For instance, while Tap-to-Pair requires at least 15 seconds, \faithful and \precise complete pairing in up to 4.1 seconds and 4.2 seconds, respectively. Note that each pairing approach requires an initialization phase, and the statistics about the initialization time are not available in many works. Therefore, we exclude the initialization time for fair comparison. However, even if the 3-second initialization time (``big silence'') is considered, the maximum pairing time of 7.1 seconds for \faithful and 7.2 seconds for \precise still shows our pairing is fast. In contrast, Perceptio~\cite{Perceptio} takes hours or even days for pairing.

\section{Usability Study}
\label{sec:user_study}

This study explores the usability, comparing with the method of inputting a password using a helper device, one of the most widely used IoT pairing mechanisms.

\begin{figure*}[!htbp]
\centering
\includegraphics[width=1\linewidth]{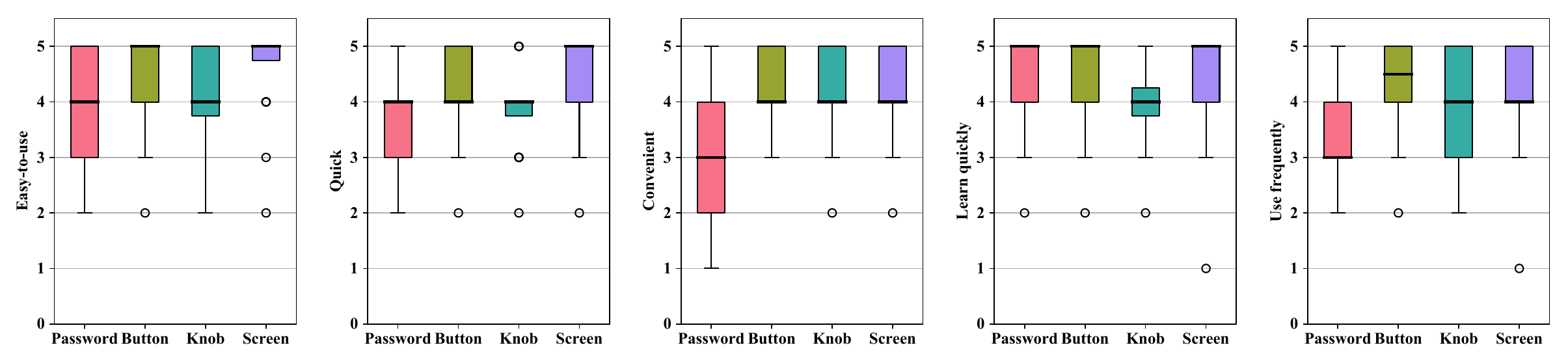}
\aaf\aaf\aaf
\caption{Usability surveyed using questions adapted from SUS~\cite{Brooke96sus}}.
\label{fig:usability}
\aaf\aaf\aaf
\end{figure*}

\subsection{Recruitment and Design}
We recruit 20 participants (9 females) by distributing recruitment flyers on the university campus. The study is advertised as ``\emph{evaluating the usability of different pairing mechanisms for IoT devices}''. Most participants are not from the Computer Science department, and none of them have a computer security background. Among the participants, 3 are local residents near the campus, 15 are students, and 2 are staff/faculty members. Their ages range from 20 to 70.

To mitigate social desirability bias, participants are \textbf{\emph{not}} informed that our system is a mechanism that we are working on. Instead, they are informed that the study aimed to investigate the usability of various pairing methods. For the password-based mechanism, following the standard Wi-Fi password protocol, which typically requires a minimum of 8 alphanumeric characters \cite{wifi_pass}, we randomly generat an 8-character alphanumeric password and show the passwaord to the participants before pairing.

The experiment is conducted in a controlled laboratory environment. Each participant is first asked to sign a consent form and complete an initial survey to provide demographic information.
To prevent learning bias, the two pairing mechanisms are introduced to participants in a random order. Specifically, our method is presented along with the three pairing operations corresponding to the three types of IoT devices, while a smartphone is used for inputting the password.
Subsequently, each participant is instructed to perform two pairing attempts on each of the three IoT devices and the smartphone to familiarize themselves with both our method and the password-based mechanism. These initial attempts are excluded from further analysis.
Following this familiarization phase, each participant performs an additional three pairing attempts on each IoT device and the smartphone, respectively.

Participants are asked to rate the following five statements on a scale from 1 to 5, where 1 indicates ``strongly disagree'', and 5 indicates ``strongly agree'': (a) I thought the pairing method was easy to use; (b) I am satisfied with the amount of time it took to complete the pairing; (c) I thought the pairing method was convenient; (d) I would imagine that most people would learn to perform the pairing very quickly; and (e) I would be happy to use this pairing method frequently. These questions are inspired by metrics utilized in previous studies \cite{bonneau2012the, usability_questionaires} and adapted from the System Usability Scale (SUS) \cite{Brooke96sus}, with adjustments made to align with our specific scenario.
Following the rating exercise, participants engage in brief interviews to provide insights into their preferences and any aspects they like or dislike about each pairing mechanism.

\aaf
\subsection{Usability Results}

\noindent\textbf{Perceived usability.}
We investigate usability across five aspects, based on the five statements mentioned above: easy to use, quick, convenient, learn quickly, and use frequently. Figure~\ref{fig:usability} presents the results. The overall scores for button clicking, knob twisting, and screen swiping are ($21.70 \pm 3.29$), ($19.80 \pm 3.76$), and ($21.65 \pm 3.54$), respectively. For password-based pairing, the overall score is ($18.45 \pm 3.37$).

To analyze the statistical significance of these results, we first hypothesize that our method exhibits similar usability compared to password-based pairing. We employ the one-way ANOVA test to examine this hypothesis. The results of the one-way ANOVA test indicate the following: (i) There are significant differences between button clicking and inputting an 8-character password ($F(1, 19)=9.057$, $p = 0.005 < 0.05$) and between screen swiping and inputting a password ($F(1, 19)=8.149$, $p = 0.007 < 0.05$). Therefore, our hypothesis can be rejected in these cases. (ii) There is no significant difference between knob twisting and inputting a password ($F(1, 19)=1.358$, $p = 0.251$). Based on the results, we conclude that users perceive better usability with button clicking and screen swiping compared to using an 8-character password, and similar usability for knob twisting and using an 8-char password.

\noindent\textbf{Pairing time.} 
We do not consider the time taken for running the pairing protocol, as our focus is solely on the time taken by the user. For our method, the mean time from the initialization to the completion of a pairing on the button, knob, and screen is $5.2\pm 0.57$s, $6.0\pm 0.83$s, and $5.6\pm 0.73$s, respectively. In contrast, for the password-based mechanism, the mean time for reading and inputting an 8-character alphanumeric password is $9.5\pm 0.78$s. Thus, our mechanism proves to be more efficient.

\noindent\textbf{Feedback.} 
We also collect their comments about the advantages and disadvantages of the pairing operations from different perspectives. Here are some representative comments: Seven subjects expressed a preference for button clicking pairing operations as they require little effort and/or burden; some also mentioned that twisting the knob for too many rounds can lead to fatigue, but they found the seven twistings used by our method to be acceptable.

\section{Related Work}
\label{sec:related_work}

\noindent\textbf{Proximity-based pairing.}
Some approaches utilize Received Signal Strength (RSS) values~\cite{Jana_2009, Bit_Extraction, Mathur_2008} or Channel State Information (CSI)~\cite{key_extraction, TDS_ccs16} to generate cryptographic keys. For instance, Move2Auth~\cite{zhang17_infocom} and SFIRE~\cite{ghose18_infocom} authenticate devices based on RSS changes correlated with smartphone motion traces. Tap-to-Pair~\cite{Tap-to-pair2018} requires a user to create RSS changes through tapping the wireless transmitter on an IoT device, adhering to instructions displayed by another device, such as a smartphone. AeroKey~\cite{aerokey} proposes to use ambient electromagnetic radiation, which exhibits spatial correlation only within a confined area, to generate symmetric keys. 
As some methods only authenticate IoT devices to the user's mobile device~\cite{zhang17_infocom, ghose18_infocom, Tap-to-pair2018}, they do not offer mutual authentication. 
This lack of mutual authentication could potentially expose IoT devices to pairing with malicious devices.
Alternatively, changes in ambient context, such as audio~\cite{audio_pairing, h2auth} and luminosity~\cite{2014CCS_CZP}, can also be used for authentication.

\zm{MPairing~\cite{mpairing} facilitates the pairing of multiple IoT devices by leveraging RSS trajectory data, which is randomized through the manual covering and uncovering of the router with aluminum foil.
Perceptio~\cite{Perceptio} clusters contextual data from diverse sensors to derive keys, offering a pairing approach applicable to various IoT devices. 
IoTCupid~\cite{iotcupid} extends Perceptio by supporting both instant and continuous sensors.
Compared to our approach, however, 
these methods rely on a physical security boundary, assuming no malicious devices exist within it. In contrast, our system significantly mitigates the threat posed by co-located malicious devices. 
Moreover, they lack assurance of correct device pairing, particularly for devices (\eg on different floors) that perceive different ambient context.

\emph{Comparison with our approach:} Unlike methods in this category, which are vulnerable to co-located malicious devices (sensing wireless signal changes or ambient context), our approach significantly mitigates this threat. In addition, in contrast to prior work that uses fuzzy commitment and suffers from information loss~\cite{mpairing, Perceptio, iotcupid}, our protocol \precise 
preserves all information for pairing.
However, unlike prior methods~\cite{mpairing, Perceptio, iotcupid} that support low-interaction or automated pairing of multiple devices, our approach requires user involvement to pair devices individually.}

\vspace{2pt}
\noindent\textbf{Physical contact-based pairing.}
Some approaches rely on physical contact or operations between users and IoT devices for pairing, but they all require special sensors (such as inertial sensors), or customization of IoT devices. For example, shaking~\cite{shakeWell} or bumping~\cite{Hinckley_2003} two devices simultaneously creates correlated motion data usable for pairing. Touch-And-Guard~\cite{touch_and_guard_tmc} involves users wearing a wristband to touch the target IoT device, utilizing the wristband's vibration motor resonance measured by accelerometers on both sides for pairing. Sethi~\etal~\cite{sync_drawing} require users to synchronize drawings on two touchscreens, with resulting drawings used for pairing. UniverSense~\cite{universense} and Posepair~\cite{posepair} both leverage a device equipped with a camera to detect the physical motion of another device equipped with an inertial sensor for device pairing. TouchKey~\cite{touchkey} utilizes the skin's electric potential induced by powerline electromagnetic radiation to pair wearable devices. By shaking \cite{Patel_2004} or moving \cite{checksum_15ubicomp} an IoT device according to a displayed public key, the key is authenticated.

\zm{\emph{Comparison with our approach:} All these methods require inertial or touch sensors embedded in the IoT device, or adding a metal pin to its surface~\cite{t2auth}, which are not available on many IoT devices. In contrast, our approach does not require any special sensors or hardware modification.}

\vspace{2pt}
\noindent\textbf{Other Approaches.} SiB \cite{seeing_is_believing} authenticates another device's public key by capturing a picture of a 2D barcode encoding the hash of the public key. VIC \cite{vic_06sp} enhances this method by presenting the key with a binary display. They require the helper device to have a camera. Plus, they do not authenticate the helper device, meaning that a malicious device can  pair with the IoT device. 

SwitchPairing~\cite{switchpairing} facilitates device pairing by connecting them to a shared power source (\ie a plug) and toggling the switch randomly. However, its reliance on a shared power source limits its applicability to IoT devices powered by plugs, excluding many devices powered by batteries, for instance. 

Many vendors embed a hard-coded password into an IoT device's firmware and print it in the user manual. This necessitates careful packaging of the device and the unique manual to ensure the correct pairing, which poses a burden to vendors \cite{ghose18_infocom, zhang17_infocom}. Some vendors use an identical password for all devices, which is a severe security flaw. Moreover, for IoT devices like a smart blood-pressure meter in Walmart that pair with many users' personal mobile devices, a single password for all users is insecure. 

\zm{\emph{Comparison with our approach:} Our approach requires no camera on the helper device, supports heterogeneous IoT devices, and does not rely on a printed password.}


\section{Discussion}
\label{sec:limitations}

\subsection{Deployment Cost}
\cxr{\precise, as well as \shortname, is designed to minimize deployment cost. It does not require any hardware modification on either the IoT device or the helper device. It operates using standard wireless communication protocols, allowing it to be integrated via firmware or software updates. \precise uses AES for commitment and decommitment operations. This is practical for modern IoT devices, as AES is widely supported across standard wireless protocols such as WiFi, Bluetooth, Zigbee, and BLE. Moreover, most commodity IoT platforms, including those based on ESP32, Nordic nRF52, and STM32 MCUs, provide either hardware-accelerated or lightweight software-based AES support.
\precise also leverages SVM, a lightweight machine learning model, to determine pairing outcomes. SVM inference is well suited for modern IoT devices, and prior work has shown that real-time SVM-based classification can be performed with minimal memory and computational overhead~\cite{ferrag2019authentication}, making it a practical solution for resource-constrained environments. Therefore, \precise is readily deployable on existing consumer devices.}

\cxr{Our system requires active user involvement and supports one-to-one pairing between a helper device and a single IoT device at a time. As the number of IoT devices in a deployment increases, this sequential pairing process may pose scalability challenges. If scalability is a primary consideration in large-scale deployments, we recommend considering alternative low-interaction or automated pairing approaches~\cite{mpairing, Perceptio, iotcupid}. However, as discussed in Section~\ref{sec:related_work}, such methods can be vulnerable to attacks by co-located malicious devices.}


\subsection{Applicability}
\cxr{As illustrated in Figure~\ref{fig:UI_survey}, our survey of the 270 most popular IoT devices on Amazon shows that only 7.4\% lack user interfaces such as buttons, knobs, or touchscreens. This demonstrates the broad applicability of our system, which can be directly applied to the remaining 92.6\% of devices without hardware modification.}

\cxr{Note that even \emph{headless} or \emph{passive} IoT devices, such as smart plugs, blinds, BLE tags, motion sensors, and contact sensors, typically include a button for pairing or reset, which can be used by our pairing approach. However, some devices, such as smart bulbs, do not have any UIs. In such cases, a user can toggle the power switch several times to turn the bulb on and off. This produces detectable state changes from which salient points can be extracted for pairing.}

\subsection{Limitations}
Our system significantly mitigates the threat due to co-located malicious devices, though it does not eliminate it entirely. If a nearby attacker has a camera pointed at the user during pairing, the system may become vulnerable to MITM 
attacks aided by computer vision techniques. However, such attacks are difficult to carry out in private spaces like homes or offices, as they require an attacker-controlled camera with a direct view of the user's pairing operations.
In public settings, we recommend users conceal their pairing actions with their body or other hand, similar to how one protects a PIN entry. If the pairing actions generate audible cues, such as button presses, a nearby device equipped with a microphone could potentially assist an MITM attack. Techniques to mitigate such side-channel attacks have been explored in prior work~\cite{anand2016sound, vibration_pairing}. 


Holding a large smartphone while twisting a small knob may not be very user-friendly. 
However, as wearable devices such as smartwatches, fitness trackers, and smart rings become increasingly popular, the usability of our system will further benefit from this trend.

\section{Conclusion}
\label{sec:conclusion}

There is an urgent need for a secure pairing approach that accommodates heterogeneous IoT devices while respecting their interface constraints. We introduce \precise{}, a secure and widely applicable pairing system for IoT devices. It requires no special sensors, hardware modifications, calibration, or clock synchronization. Through simple physical operations, users can conveniently complete the pairing process within a few seconds. 
In addition, we propose a pairing protocol, \pname{}, that achieves zero information loss, eliminating the need for evidence encoding and enabling higher accuracy. A comprehensive evaluation, including a user study, demonstrates the high security, usability, stability, and efficiency of our approach.

\section{Acknowledgment}
\label{sec:acknowledgment}

\crm{This work was supported in part by the US National Science Foundation (NSF) under grants CNS-2309550 (CAREER Award). This work was also supported in part by the Commonwealth Cyber Initiative (CCI) in Virginia. The authors would like to thank the anonymous reviewers for their valuable comments and constructive suggestions. }
\bibliographystyle{IEEEtran}
\bibliography{sections/reference.bib}


\vspace{-22pt}
\begin{IEEEbiography}[{\includegraphics[width=1in,height=1.25in,clip,keepaspectratio]{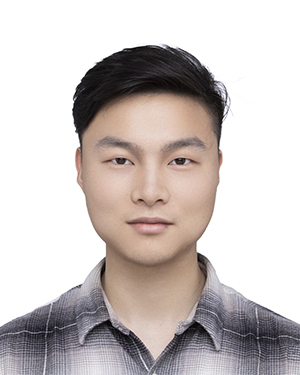}}]{Chuxiong Wu} is an Assistant Professor at Southern Illinois University. He received his BE degree in Electronic Information Engineering from Beihang University, MS degree in Computer Science from the University of South Carolina, and PhD degree in Computer Science from George Mason University. His research interest focuses on Security in Cyber-Physical Systems.
\end{IEEEbiography}

\vspace{-22pt}
\begin{IEEEbiography}
[{\includegraphics[width=1in,height=1.25in,clip,keepaspectratio]{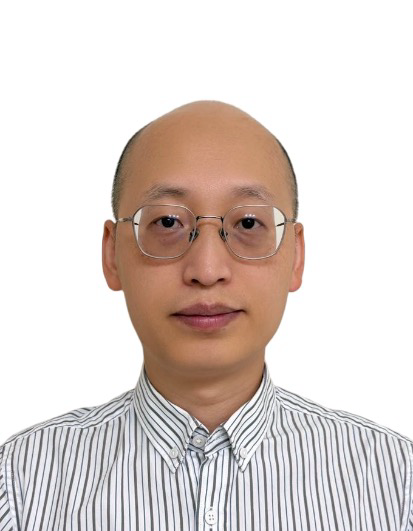}}]{Xiaopeng Li} received his PhD in Computer Science from the University of South Carolina, MS in Instrument Science and Technology from Southeast University, and BS in Measurement and Control Technology from Harbin Institute of Technology. He is an Assistant Professor of Computer Science at the University of Central Oklahoma and formerly worked at Microsoft. His research focuses on system security, AI-driven security, and the security and privacy of AI systems.
\end{IEEEbiography}

\vspace{-22pt}
\begin{IEEEbiography}
[{\includegraphics[width=1in,height=1.25in,clip,keepaspectratio]{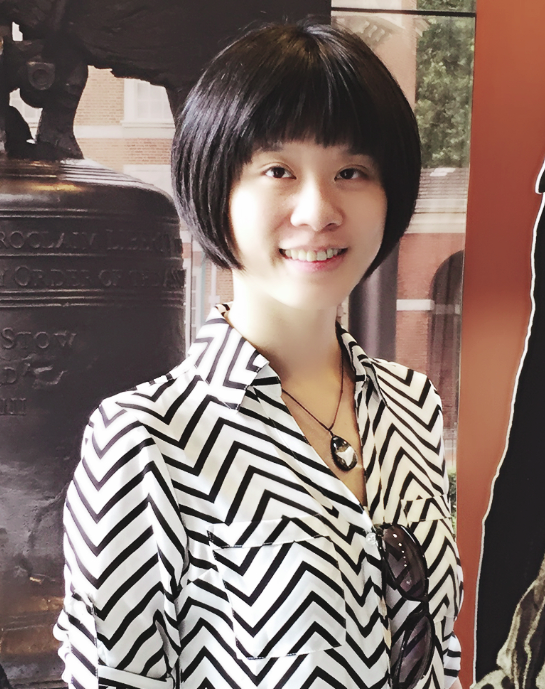}}]{Lannan Luo}
is an Associate Professor at George Mason University. She received
received her BS degree from Xidian University in 2009, MS degree from the 
University of Electronic Science and Technology of China in 2012, and PhD 
degree from the Pennsylvania State University in 2017. 
Her research interests are software and system security.
\end{IEEEbiography}

\vspace{-22pt}

\begin{IEEEbiography}[{\includegraphics[width=1in,height=1.25in,clip,keepaspectratio]{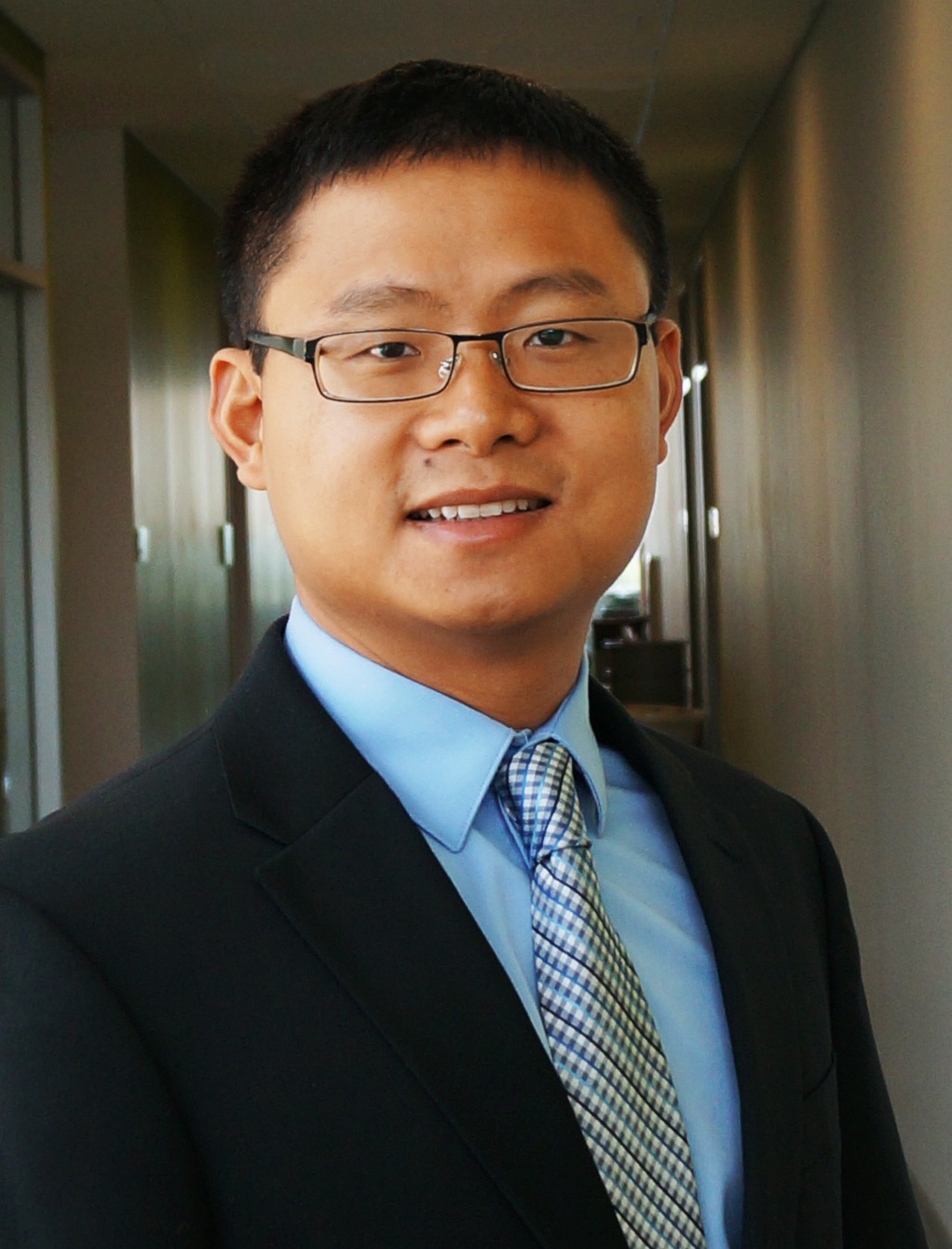}}]{Qiang Zeng}
is an Associate Professor in the Department of Computer Science at George Mason University.
He received the bachelor’s and master’s degrees from Beihang University, and PhD degree from Penn State University. He is the recipient of an NSF CAREER Award. His main research interest is Computer Systems Security, with a focus on Cyber-Physical Systems and Internet of Things. 
\end{IEEEbiography}

\newpage
\section{Sensing Pairing Operations (Buttons and Screens)}
\label{sec:appendix_a}

Correlation of IMU data and pairing operations.

\begin{figure*}[hbt!]
\begin{minipage}{1\textwidth}
  \centering
\includegraphics[scale=0.29]{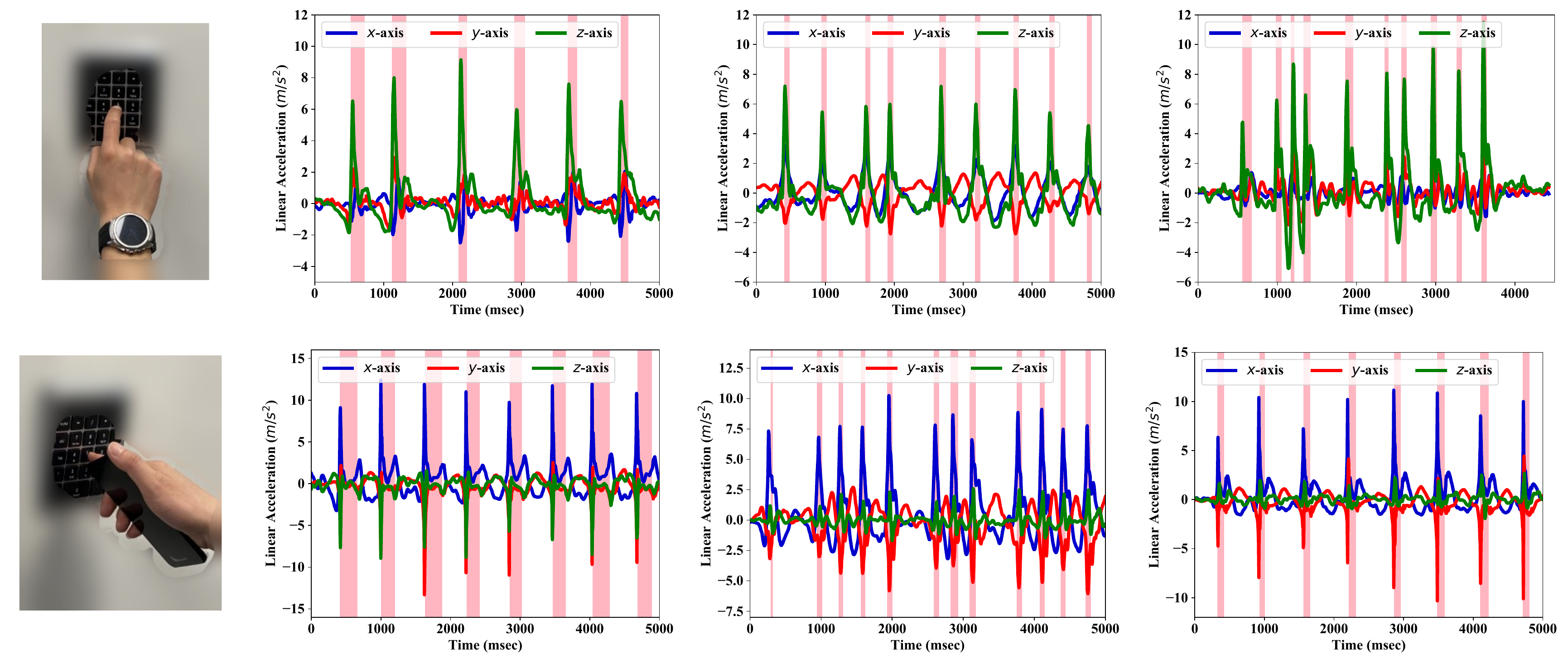} \\
\begin{tabular}{P{3.5cm} P{4.0cm} P{4.4cm} P{4.5cm}}
Demonstration & User 1 & User 2 & User 3
\end{tabular}
\caption{The acceleration data captured when
users press buttons, and their correlation with button-pressing operations.}
\label{fig:motion_data_button}
\end{minipage}

\vspace{30pt}

\begin{minipage}{1\textwidth}
  \centering
\includegraphics[scale=0.29]{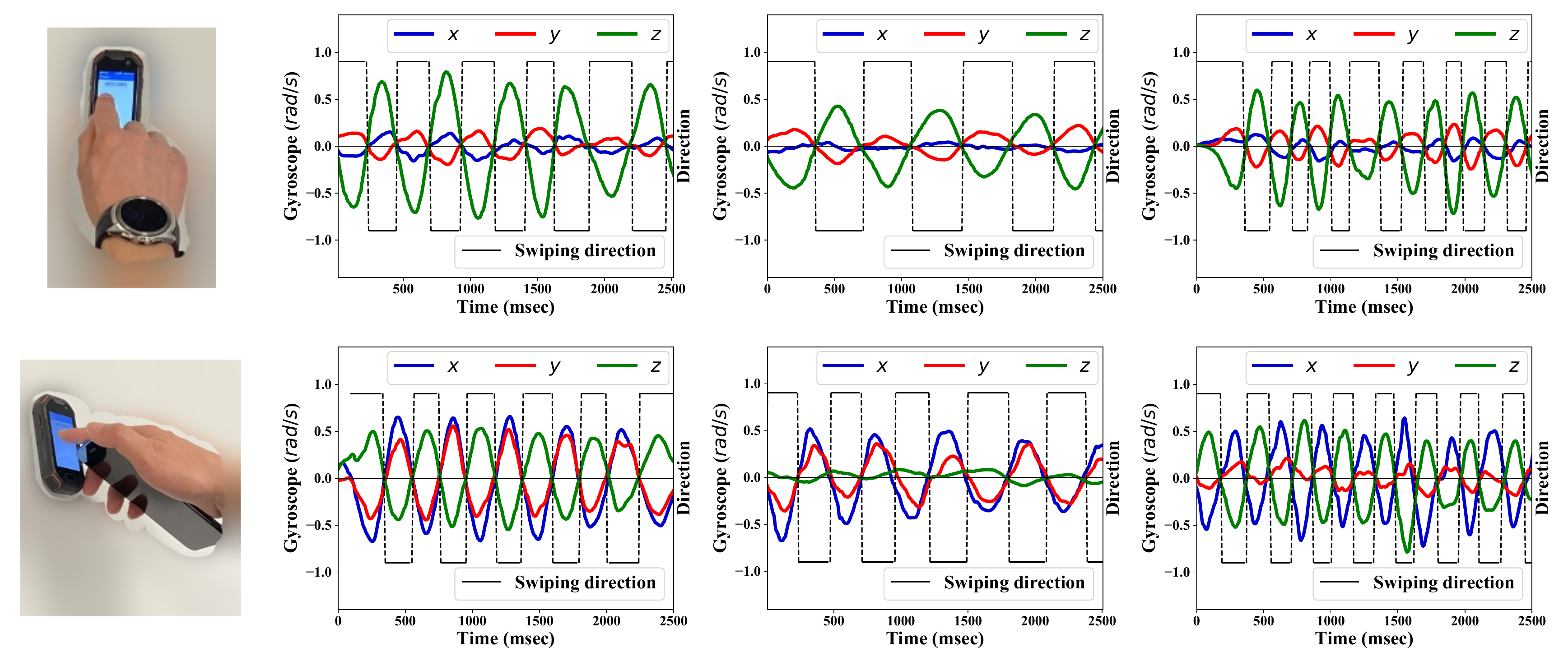} \\
\begin{tabular}{P{3.5cm} P{4.2cm} P{4.4cm} P{4.5cm}}
Demonstration & User 1 &  User 2 & User 3
\end{tabular}
\caption{The gyroscope data captured when
users swipe touchscreens, and their correlation with swiping operations.}
\label{fig:motion_data_screen}
\end{minipage}

\end{figure*}
\end{document}